\documentclass[sigconf,screen]{acmart}
\newcommand{\tool}{\textsc{MARL-OT}}
\usepackage{multirow}
\usepackage{color}
\usepackage{subcaption}
\usepackage{algorithm,algorithmic}
\usepackage{amsmath}
\usepackage{multicol}
\begin{document}

\title{MARL-OT: Multi-Agent Reinforcement Learning Guided Online Fuzzing to Detect Safety Violation in Autonomous Driving Systems}


\author{Linfeng Liang}
\affiliation{%
  \institution{Macquarie University}
  \city{Sydney}
  \state{NSW}
  \country{Australia}}
\email{linfeng.liang@hdr.mq.edu.au}

\author{Xi Zheng}
\affiliation{%
  \institution{Macquarie University}
  \city{Sydney}
  \state{NSW}
  \country{Australia}}
\email{James.Zheng@mq.edu.au}


\begin{abstract}

Autonomous Driving Systems (ADSs) are safety-critical, as real-world safety violations can result in significant losses. Rigorous testing is essential before deployment, with simulation testing playing a key role. However, ADSs are typically complex, consisting of multiple modules such as perception and planning, or well-trained end-to-end autonomous driving systems. Offline methods, such as the Genetic Algorithm (GA), can only generate predefined trajectories for dynamics, which struggle to cause safety violations for ADSs rapidly and efficiently in different scenarios due to their evolutionary nature. Online methods, such as single-agent reinforcement learning (RL), can quickly adjust the dynamics' trajectory online to adapt to different scenarios, but they struggle to capture complex corner cases of ADS arising from the intricate interplay among multiple vehicles. Multi-agent reinforcement learning (MARL) has a strong ability in cooperative tasks. On the other hand, it faces its own challenges, particularly with convergence. This paper introduces \tool, a scalable framework that leverages MARL to detect safety violations of ADS resulting from surrounding vehicles' cooperation. \tool\ employs MARL for high-level guidance, triggering various dangerous scenarios for the rule-based online fuzzer to explore potential safety violations of ADS, thereby generating dynamic, realistic safety violation scenarios. Our approach improves the detected safety violation rate by up to 136.2\% compared to the state-of-the-art (SOTA) testing technique.

\end{abstract}


% \keywords{Autonomous driving system, Multi-agent reinforcement learning, Search-based testing, Online testing}


\maketitle

\section{Introduction}
\begin{figure}[H]
    \centering
    \includegraphics[width=1\linewidth]{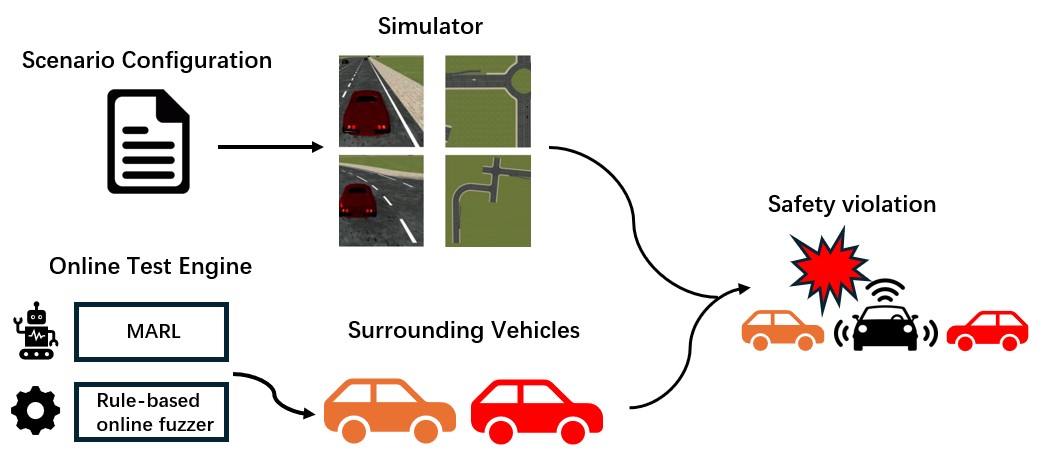}
\caption{A high-level overview of \tool \ framework. }
\label{overall}  
\end{figure}

Autonomous Driving Systems (ADSs) are complex, typically consisting of multiple modules working together \cite{ap, tesla_autopilot} or well-trained end-to-end autonomous driving systems \cite{chen2024end} to ensure reliability and safety. Given the critical importance of ADS safety \cite{bertoncello2015ten}, thorough evaluation is essential before real-world deployment \cite{barr2014oracle}. However, real-world testing demands vast driving miles, leading to high costs and time commitments \cite{lambert2016understanding}. In contrast, simulation-based testing offers benefits such as controlled environments, the ability to test diverse scenarios, and the creation of dangerous situations without endangering human lives. As a result, researchers have developed various methods for constructing virtual scenarios to assess ADSs \cite{liang2023rlaga, deng2022scenario}.

Simulation testing is widely used for ADSs \cite{li2020av, deng2022scenario, haq2022many, lou2022testing}, with recent studies showing that scenario-based testing can generate critical corner cases through both online and offline approaches \cite{tian2022mosat, li2020av, liang2023rlaga, feng2023dense}. Offline approaches, such as GA, rely on evolutionary processes across generations to generate predefined violation configurations through continuous searching. However, GA struggles to effectively generate trajectories for dynamics, like throttle and steering sequences for vehicles, especially in scenarios with multiple dynamics \cite{tian2022mosat, li2020av}. This limitation hampers the exploration of the search space, leading to missed critical corner cases \cite{liang2023rlaga} (\emph{challenge 1}).

Recent research indicates that online methods, such as RL, are effective for real-time scenario adaptation and corner case generation \cite{feng2023dense, koren2021finding, liang2023rlaga}. Techniques like backward training \cite{koren2021finding} and surrogate training \cite{liang2023rlaga} have been developed to enhance RL agents' convergence. However, in ADS testing context, single-agent systems struggle to detect safety violations resulting from interactions between multiple surrounding vehicles, which are common in real-world accidents (\emph{challenge 2}). Meanwhile, MARL faces significant convergence challenges, even with rapid convergence techniques, due to the large action and state spaces in simulations and the rarity of violations (\emph{challenge 3}).

This paper introduces a novel method, Multi-Agent RL Online Testing (\tool), to address these challenges. Figure \ref{overall} provides an overview of \tool: within a given scenario, \tool \ uses a pre-trained MARL system for high-level guidance, which converges in 15 minutes. The rule-based fuzzing logic then maps the MARL output to natural driving behavior, triggering various dangerous scenarios for the rule-based online fuzzer to manipulate. This generates natural, realistic driving behaviors for surrounding vehicles and creates cooperative scenarios that lead to safety violations by the ego vehicle. This approach establishes an efficient, scalable pipeline for generating safety violations in ADS. Our approach improves the detected safety violation rate by up to 136.2\% compared to the SOTA testing technique. Our contributions are as follows:

\begin{itemize} 
\item We propose \tool, an adaptive MARL-based online testing tool that detects ego vehicle violations driven by cooperative behaviors among surrounding vehicles.
\item We propose a rule-based online fuzzing approach that models driving maneuvers by mapping continuous MARL agent outputs to discrete actions, triggering realistic and smooth rule-based patterns for each surrounding vehicle.

%\tool \ can detect normal, natural, and realistic violations led by the ego vehicle controlled by ADSs.

\item 
We conduct extensive experiments across different scenarios and ADSs to demonstrate that our method effectively detects normal, natural, and realistic violations caused by the ego vehicle controlled by ADSs.

%and the realism of the detected violations.
\end{itemize}

This paper is organized as follows: In Section \ref{RW}, we provide a review and discussion of the related work relevant to our study. Section \ref{method} details the development of our method. In Section \ref{exp}, we describe the design of our experiments. Section \ref{result} presents the results of our comprehensive experiments based on the methodology outlined in Section \ref{method}. In Section \ref{dis}, we discuss the utility and the future work related to our proposed method. Conclusions are summarized in Section \ref{conclusion}.

\section{Related Work}
\label{RW}
% \subsection{Multi-Agent Reinforcement Learning}

\subsection{
Search-Based Offline Testing
%Violation Detection in Cyber-Physical System
}
\par 
The efficacy of GA in identifying corner test cases that trigger violations and failures in cyber-physical systems has been well-documented \cite{li2020av, schmidt2022stellauav, panichella2015reformulating, abdessalem2018testing, ben2016testing, ebadi2021efficient, huai2023sceno, luo2021targeting}. GA is widely used in search-based software testing, typically as an offline search technique \cite{wegener2004evaluation}.

In \cite{abdessalem2018testing, ben2016testing}, new objective functions with multiple test goals were proposed to guide the search for test scenarios to evaluate Advanced Driver Assistance Systems (ADAS). In \cite{ebadi2021efficient}, a search-based method was introduced to test Baidu Apollo's pedestrian detection algorithm \cite{Apollo}, manipulating static parameters like weather and the positions of dynamic objects. AV-FUZZER \cite{li2020av} combines global and local fuzzers based on GA to find corner cases in ADSs. AutoFuzz \cite{zhong2022neural} shows that neural networks can enhance GA by using gradients to mutate seeds. ScenoRITA \cite{huai2023sceno} proposes new gene representations for testing scenarios, allowing obstacles to be fully mutable and improving violation detection. Recent work, MOSAT \cite{tian2022mosat}, uses GA to manipulate driving maneuver patterns from real-world crash scenarios to better identify violations in ADSs.

These works focus on innovative scenario representations, objective functions, and crossover and mutation operations, generating test scenarios offline. However, due to their evolutionary nature, offline methods struggle to quickly generate action sequences, such as throttle and steering, for ADS, particularly when handling multiple scenarios. Additionally, they limit interactions between dynamic objects and the ego system, reducing the discovery of corner cases. In contrast, we propose an online fuzzer guided by MARL to search dynamic actions online, maximizing the likelihood of rapidly identifying safety violations in a given scenario.
% a specially designed GA with targeted mutations and crossover operations to enhance the generation of diverse online search spaces for RL. Our approach uses RL to dynamically alter the complex interactions and trajectories of dynamic objects.
\subsection{RL-based Online Testing}

The effectiveness of RL in generating violation cases for software testing has been well-demonstrated \cite{lu2022rgchaser, koren2018adaptive}. RL operates as a Markov Decision Process (MDP), involving agents, actions, policies, and rewards in an interactive environment \cite{sutton2018reinforcement}. In an MDP, the agent perceives the current state, takes actions based on a policy, and receives rewards. RL can be used for online testing to dynamically generate corner cases through predefined reward functions \cite{koren2018adaptive, lu2022learning}. Research indicates that RL-based online testing approaches can control dynamic objects in real-time within test scenarios, exploring a broader range of potential violations \cite{koren2018adaptive, lu2022learning}. D2RL introduces an RL-based method to fully capture the ADS testing environment, though it requires significant computational resources and long training periods \cite{feng2023dense}. GARL \cite{liang2023rlaga} combines GA and RL to generate diverse scenarios and dynamic trajectories that trigger violations in UAV landing scenarios.

MARL has been widely deployed for both competitive and cooperative tasks, such as in StarCraft \cite{vinyals2019grandmaster} and UAV target capture tasks \cite{jiang2023uavs}. Compared to single-agent RL, MARL more closely aligns with real-world scenarios, where many accidents result from interactions between multiple vehicles, which can be viewed as cooperative interactions. However, in simulation-based testing, due to the vast action and state spaces and the rarity of violations, MARL faces significant convergence problems. Although previous methods such as backward training \cite{koren2021finding} and surrogate environment training \cite{liang2023rlaga} have been proposed to address the convergence problem, it still remains a challenge to map actions from the simplified training environment to the full simulation in a way that naturally results in real and common safety violations.

To address this, we create simplified scenarios to train MARL agents and apply rule-based fuzzing logic to map the MARL output to driving behavior. We then use a rule-based online fuzzer, triggered by MARL, to generate natural, real-world driving maneuvers that cause safety violations by the ego vehicle.

 % the full simulation. Our framework combines an online fuzzer with road condition constraints to regulate NPC vehicle behavior and maximize violations led by the ego vehicle.

% A backward training strategy \cite{koren2021finding} has been proposed to mitigate this by using expert demonstrations in low-fidelity simulations to train the agent before deployment in high-fidelity simulations, which can cause transferability issues. Unlike existing RL-based methods, our framework uniquely combines GA and RL, using GA to simplify RL's learning dimensions. Our surrogate training eliminates the need for expert demonstrations, enhancing agent transferability and allowing direct application in high-fidelity simulations.
% \subsection{LLM-involved Testing}

% \par Recent works indicates that LLMs could be possible to be used to benefit the GA
% Recent works suggest that LLMs can be used as backbones for performing downstream tasks, such as semantic segmentation \cite{lai2024lisa} and robotic arm actions \cite{shentu2024llms}.
\section{Method}
\label{method}

\begin{figure*}[ht]
    \centering
    \includegraphics[width=0.9\linewidth]{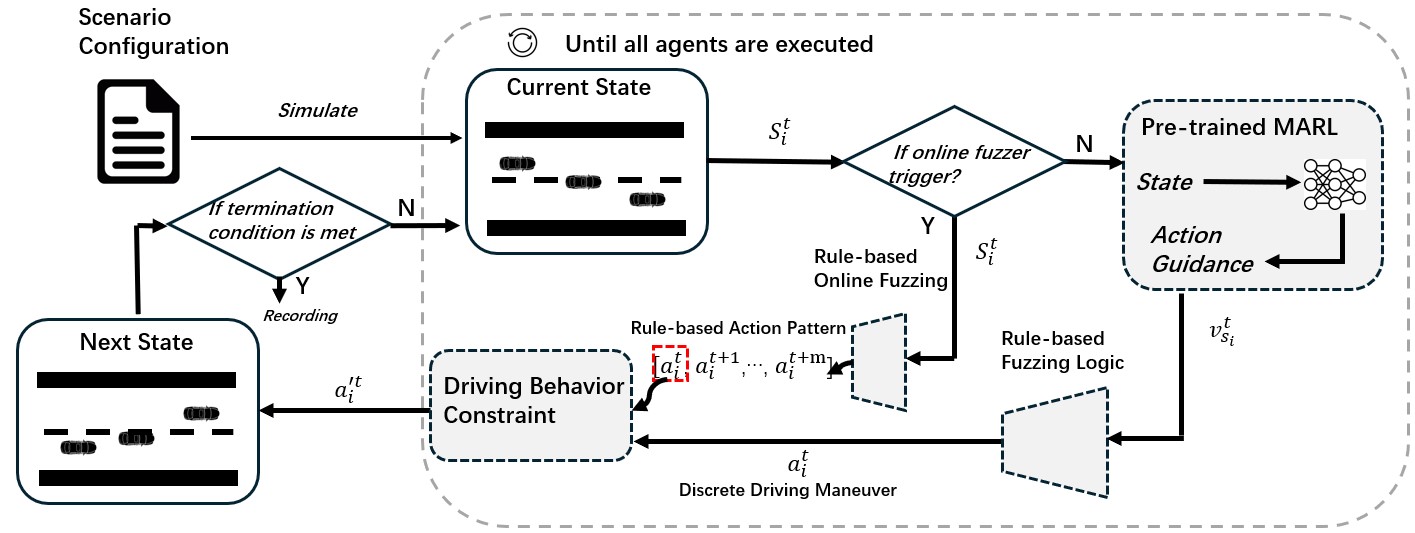}
	\caption{Workflow of \tool.}
\label{workflow}  
\end{figure*}
\subsection{Overview}

Figure \ref{workflow} illustrates the workflow of the \tool\ framework. First, within an initialized scenario in the simulator, the current state of each agent is evaluated to determine if it triggers the online fuzzer. If \textit{True}, the online fuzzer selects the appropriate rule-based action pattern that meets the triggered condition as the next action of the surrounding vehicle, continuing until the pattern is completed. If \textit{False}, the state is sent to the pre-trained MARL to determine the agent’s movement vector for the next time step. The movement vector is then mapped to a discrete driving maneuver of the surrounding vehicle through the fuzzing logic, and both the rule-based action pattern and the discrete driving maneuver are regulated by driving behavior constraints to ensure safe driving behavior. This loop continues until all surrounding vehicles have executed their actions. The test ends when the termination condition is met, i.e., the ego vehicle reaches its destination safely or a safety violation occurs. If a safety violation occurs, the case is recorded.

\subsection{Multi-Agent Reinforcement Learning}

In our task, MARL operates with a cooperative objective, where surrounding vehicles aim to keep their positions around the ego vehicle, allowing the online fuzzer to manipulate the surrounding vehicles within the predefined rule-based action pattern. MADDPG \cite{lowe2017multi} has demonstrated strong performance in both performance and convergence, making it an ideal choice for this context.

\subsubsection{State}
The state (\( S \)) vector encodes the full observation of the environment, which includes positional and motion data for each surrounding vehicle and the ego vehicle. For agent \( i \), the state is:

\begin{equation}
S^t_i = (P^t_{s_i,x}, P^t_{s_i,y}, v^t_{s_i,x}, v^t_{s_i,y}, \ldots, P^t_{s_{(i+n)},x}, P^t_{s_{(i+n)},y}, P^t_{e,x}, P^t_{e,y})
\end{equation}

where:
\begin{itemize}
    \item \( P^t_{s_i,x} \) and \( P^t_{s_i,y} \): \( i \)-th surrounding vehicle's position at $t$-th time step,
    \item $v^t_{s_i,x}$ and $v^t_{s_i,y}$: the $i$-th surrounding vehicle's length of movement in the $x$ and $y$ components at the $t$-th time step, respectively. These are initialized with \([0,0]\).

    \item \( P^t_{e,x} \) and \( P^t_{e,y} \): ego vehicle's position at $t$-th time step,
    \item $n$: the number of available surrounding vehicles.
\end{itemize}

\subsubsection{Action Space}
Each surrounding vehicle's action ($A$) at $t$-th time step is a continuous 2-tuple vector denote as:

\begin{equation}
    A^t_i := \{\Delta v^{t}_{s_i,x}, \Delta v^{t}_{s_i,y}\}
\end{equation}

where:
\begin{itemize}
    \item $\Delta v^{t}_{s_i,x}$: the change of the x-axis movement length at $t$-th time step, with a range of \([-0.1, 0.1]\),
    \item $ \Delta v^{t}_{s_i,y}$: the change of the y-axis movement length at $t$-th time step, with a range of \([0, 0.1]\).%\jz{define horizontal and vertical. Magnitude is horizontal and direction is vertical? Also magnitude means acceleration? direction means angular direction? what does -0.1 and 0.1 means for magnitude and what does -0.1 and 0.1 means for direction?}
\end{itemize}

We enforce $ \Delta v_{s_i,y}>0$ in our training environment, as vehicles are restricted to forward movement only. %\jz{bigger than 0 means move forward, smaller than 0 means move backforward, so direction is forward and backward? how about left and right? Or you actually means horizontal is left and right and vertical forward and backward? If so, what happen to magnitude? be careful of your wording and consistency with the formula, these reviewers are system people and many of them are formal methods people, they are very strong in formula and maths. Computer science has two foundational categories relying heavily on maths, one is of course machine learning, another is formal methods.}

\subsubsection{Reward}
We define a three-phase reward function that encourages cooperative behavior for surrounding vehicles: the first phase incentivizes surrounding vehicles to initially approach the ego vehicle, the second phase promotes encircling the ego vehicle, and the third phase rewards a full enclosure around the ego vehicle. The reward function is as follows:
% \jz{you need to define quickly what exactly is three-phase reward function, i.e., initial, intermediate, final and explain this suits for our scenario which is complex, multi-step behavior of surrounding vechiles, once again removing NPC vehicles }
\subsubsection*{Phase 1: Proximity Reward ($r_{\text{near}}$)}
surrounding vehicles are rewarded for moving closer to the ego vehicle from their initial positions, with alignment of their movement vector toward the ego vehicle:

\begin{equation}
v_{s_i}^t = (v_{s_i, x}^t, v_{s_i, y}^t)
\end{equation}
\begin{equation}
P_{s_i}^t = (P_{s_i, x}^t, P_{s_i, y}^t)
\end{equation}
\begin{equation}
P_e^t = (P_{e, x}^t, P_{e, y}^t)
\end{equation}
\begin{equation}
    \cos(\theta_i^t) = \frac{v_{s_i}^t \cdot (P_{s_i}^t - P_e^t)}{\| v_{s_i}^t \|_2 \| P_{s_i}^t - P_e^t \|_2 + 0.001}
\end{equation}
% \begin{equation}
% \cos(\theta_i^t) = \frac{v_{s_i, x}^t (P_{s_i, x}^t - P_{e, x}^t) + v_{s_i, y}^t (P_{s_i, y}^t - P_{e, y}^t)}{\sqrt{v_{s_i, x}^t{}^2 + v_{s_i, y}^t{}^2} \cdot \sqrt{(P_{s_i, x}^t - P_{e, x}^t)^2 + (P_{s_i, y}^t - P_{e, y}^t)^2} + 0.001}
% \end{equation}
\begin{equation}
    r^t_{i,\text{near}} = \| v_{s_i}^t \|_2 \cos(\theta_i^t)
\end{equation}

% \jz{in your previous defintion, the positon contains two scalar one is x and another is y, be consistent. similar to vi, which has horizontal and vertical component, maybe explain as x and y instead. You basically say you want do vector level operation, each vector contains x and y scalar as explained before. Such a sentence shall be sufficient. You give vector form for vi but not for Pn and Pe yet, be regid}

% where:
% \begin{itemize}
%     \item \( v_i \): i-th NPC vehicle's vector,
    % \item \( v_{\text{max}} \): speed cap,
    % \item \( \cos(\theta_i) \): cosine of the angle between the velocity vector and the position difference vector from the i-th NPC vehicle to the ego-vehicle.
% \end{itemize}
% This reward encourages NPC vehicles to align their movement toward the ego vehicle, rewarding those that get closer to the ego vehicle while maintaining alignment.
\subsubsection*{Phase 2: Encirclement Reward}
The surrounding vehicle transitions from proximity to encirclement. Given the complexity of encirclement, we break it down into three stages. The first stage involves the surrounding vehicle tracking the ego vehicle, the second stage involves the surrounding vehicle encircling the ego vehicle, and the third stage involves the surrounding vehicle achieving full enclosure of the ego vehicle. The reward functions are denoted as:

\subsubsection*{Stage 1: Tracking Reward ($r_{\text{track}}$)}
Surrounding vehicles are rewarded for staying close and beginning to encircle the ego vehicle.
% \begin{equation}
% \begin{aligned}
% Area^t_i = \frac{1}{2} \left| P_{e, x}^t (P_{s_i, y}^t - P_{s_{i+1}, y}^t) + P_{s_i, x}^t (P_{s_{i+1}, y}^t - P_{e, y}^t) \right. \\
% \left. + P_{s_{i+1}, x}^t (P_{e, y}^t - P_{s_i, y}^t) \right|
% \end{aligned}
% \end{equation}

\begin{equation}
    \text{if} \quad 
    \begin{aligned}
        & \sum_{i=1}^{n} Area^t_i > Area^t_t, \\
        % & \sum_{i=1}^{n} \| P_{s_i}^t - P_e^t \|_2 \geq d_{\text{limit}}, \\
        & \min_{i} \| P_{s_i}^t - P_e^t \|_2 \geq d_{\text{enclosure}}
    \end{aligned}
\end{equation}

\begin{equation}
r^t_{\text{track}} = 
\begin{cases} 
    -\frac{\sum_{i=1}^{n} \| P_{s_i}^t - P_e^t \|_2}{\max_{i=1}^{n} \| P_{s_i}^t - P_e^t \|_2} 
    & \text{if conditions above hold}, \\
    0 & \text{otherwise}
\end{cases}
\end{equation}

% \begin{equation}
% r^t_{\text{track}} = 
% \begin{cases} 
%     -\frac{\sum_{i=1}^{n} \| P_{s_i}^t - P_e^t \|_2}{\max_{i=1}^{n} \| P_{s_i}^t - P_e^t \|_2} 
%     & \text{if} \quad
%     \begin{aligned}
%         & \sum_{i=1}^{n} Area^t_i > Area^t_t, \\
%         & \sum_{i=1}^{n} \| P_{s_i}^t - P_e^t \|_2 \geq d_{\text{limit}}, \\
%         & \min_{i} \| P_{s_i}^t - P_e^t \|_2 \geq d_{\text{enclosure}}
%     \end{aligned} \\
%     0 & \text{otherwise}
% \end{cases}
% \end{equation}
% \begin{equation}
% r^t_{\text{track}} = 
% \begin{cases} 
%     -\frac{\sum_{i=1}^{n} \| P_{s_i}^t - P_e^t \|_2}{\max_{i=1}^{n} \| P_{s_i}^t - P_e^t \|_2} 
%     & \text{if } \sum_{i=1}^{n} Area^t_i  > Area^t_t \\
%     & \quad \text{and } \sum_{i=1}^{n} \| P_{s_i}^t - P_e^t \|_2 \geq d_{\text{limit}} \\
%     & \quad \text{and } \| P_{s_i}^t - P_e^t \|_2 \geq d_{\text{enclosure}} \\
%     & \quad \text{ for all } i = 1, 2, \dots, n \\
%     0 & \text{otherwise}
% \end{cases}
% \end{equation}

% \begin{equation}
% r_{\text{track}} = 
% \begin{cases} 
%     -\frac{\sum_{i=1}^{n} \| P_{s_i}^t - P_e^t \|_2}{\max_{i=1}^{n} \| P_{s_i}^t - P_e^t \|_2} 
%     & \text{otherwise}
% \end{cases}
% \end{equation}
% \begin{aligned}
% \text{subject to } & \quad \sum_{i=1}^{n} S_i  > S_t, \\
%                    & \quad \sum_{i=1}^{n} \| P_{s_i}^t - P_e^t \|_2 \geq d_{\text{limit}}, \\
%                    & \quad \| P_{s_i}^t - P_e^t \|_2 \geq d_{\text{enclosure}} \text{ for all } i = 1, 2, \dots, n
% \end{aligned}

\subsubsection*{Stage 2: Encircling Reward ($r_{\text{encircle}}$)}
As surrounding vehicles form a partial enclosure around the ego vehicle, they receive an additional reward.
% \begin{equation}
% r^t_{\text{encircle}} = 
% \begin{cases} 
%     -\frac{\log\left( \sum_{i=1}^{n} Area^t_i - Area^t_t + 1 \right)}{n} 
%     & \text{if} \quad
%     \begin{aligned}
%         & \sum_{i=1}^{n} Area^t_i > Area^t_t, \\
%         & \exists i, \quad \| P_{s_i}^t - P_e^t \|_2 \in \left[ d_{\text{limit}}, d_{\text{enclosure}} \right]
%     \end{aligned} \\
%     0 & \text{otherwise}
% \end{cases}
% \end{equation}
% \begin{equation}
% r^t_{\text{encircle}} = 
% \begin{cases} 
%     -\frac{\log(\sum_{i=1}^{n} Area^t_i - Area^t_t + 1)}{n} 
%     & \text{if } \sum_{i=1}^{n} Area^t_i  > Area^t_t \\
%     & \quad \text{and } (\sum_{i=1}^{n} \| P_{s_i}^t - P_e^t \|_2 < d_{\text{limit}} \\
%     & \quad \text{or } \| P_{s_i}^t - P_e^t \|_2 \geq d_{\text{enclosure}} \\
%     & \quad \exists i = 1, 2, \dots, n) \\
%     0 & \text{otherwise}
% \end{cases}
% \end{equation}

% \begin{equation}
%    r^t_{\text{encircle}} = -\frac{\log(\sum_{i=1}^{n} Area^t_i - Area^t_t + 1)}{n} 
% \end{equation} 
\begin{equation}
    \text{if} \quad 
    \begin{aligned}
        & \sum_{i=1}^{n} Area^t_i > Area^t_t, \\
        & \exists i = \{1, 2, \dots, n\} ,\| P_{s_i}^t - P_e^t \|_2 < d_{\text{enclosure}} 
    \end{aligned}
\end{equation}
\begin{equation}
r^t_{\text{encircle}} = 
\begin{cases} 
    -\frac{\log(\sum_{i=1}^{n} Area^t_i - Area^t_t + 1)}{n} 
    & \text{if the conditions above hold}, \\
    0 & \text{otherwise}
\end{cases}
\end{equation}

\subsubsection*{Stage 3: Full Enclosure Reward ($r_{\text{full}}$)}
A full enclosure around the ego vehicle yields a reward:

\begin{equation}
    \text{if} \quad 
    \begin{aligned}
        & \sum_{i=1}^{n} Area^t_i = Area^t_t, \\
        & \exists i = \{1, 2, \dots, n\} ,\| P_{s_i}^t - P_e^t \|_2 > d_{\text{enclosure}} 
    \end{aligned}
\end{equation}

\begin{equation}
\hspace{-0.5cm}r^t_{\text{full}} = 
\begin{cases} 
     \exp\left(\frac{\sum_{i=1}^{n} \| P_{s_i}^t - P_e^t \|_2 - \sum_{i=1}^{n} \| P_{s_i}^{t+1} - P_e^{t+1} \|_2}{n}\right) 
    & \text{if conditions above hold}, \\
    0 & \text{otherwise}
\end{cases}
\end{equation}

% \begin{equation}
%    r^t_{\text{full}} =  \exp\left(\frac{\sum_{i=1}^{n} \| P_{s_i}^t - P_e^t \|_2 - \sum_{i=1}^{n} \| P_{s_i}^{t+1} - P_e^{t+1} \|_2}{n}\right)
% \end{equation}

where:
\begin{itemize}
    % \item \( d_i \): distance of i-th surrounding vehicle to the ego vehicle,7
    % \item \( d^`_i \): distance of i-th surrounding vehicle to the ego vehicle at last timestep,
    % \item \(d_{\text{limit}}\): limitation on the distance required to initiate encirclement.
    \item \(d_{\text{enclosure }}\): limitation on the distance required to initiate enclosure.
    % \item \( \sum_{i=1}^{n} S_i \): total area of triangles formed by $i$-th and $i+1$-th NPC vehicles and ego vehicle,\jz{this does not sound right, what happ to N-th vechile, where is N+1 surrounding vehicle?}
    \item $Area^t_i$: the area of the triangle formed by the $i$-th surrounding vehicle, its adjacent surrounding vehicle, and the ego vehicle.
    \item \( Area^t_t \): the total area encompassing all surrounding vehicles. %\jz{what do you mean area within all surrounding vecicles? Do you mean the total area encompassing all surrounding vehicles? How do you define this area?}.
\end{itemize}

\subsubsection*{Phase 3: Completion Reward ($r_{\text{finish}}$)}
Upon successfully encircling the ego vehicle, surrounding vehicles receive a substantial reward:

\begin{equation}
r^t_{\text{finish}} = 
\begin{cases} 
     \mathbf{I} 
    & \text{if} \quad
    \begin{aligned}
        & \sum_{i=1}^{n} Area^t_i  = Area^t_t, \\
        & \max_{i} \| P_{s_i}^t - P_e^t \|_2 \leq d_{\text{enclosure}}
    \end{aligned} \\
    0 & \text{otherwise}
\end{cases}
\end{equation}

This final reward reinforces mission completion and concludes the episode, where $\mathbf{I}$ is a completion indicator. The overall reward function of each surrounding vehicle at time step $t$ can be denoted as:
\begin{equation}
    \mathbf{R}^t_i = \mu_1 r^t_{i,\text{near}} + \mu_2 ( r^t_{\text{track}}+ r^t_{\text{encircle}} + r^t_{\text{full}}) +\mu_3 r^t_{\text{finish}}
\end{equation}
where $\mu_1$ to $\mu3$ are coefficients for each reward component, set empirically. 

Additionally, beyond the cooperation reward among agents, we also establish a competition reward for the ego vehicle to encourage it to evade the surrounding vehicle. This reward can be expressed as:
\begin{equation}
    r^t_{\text{ego}} = \sum_{i=1}^{n} \| P_{s_i}^{t+1} - P_e^{t+1} \|_2 - \sum_{i=1}^{n} \| P_{s_i}^t - P_e^t \|_2 
\end{equation}

\subsubsection{Training}
Initially, we trained the MARL agents in our customized OpenAI Gym environment \cite{brockman2016openaigym}. In this environment, the surrounding and ego vehicles can move freely along their output actions without being constrained by road networks. The framework, however, is adaptable and does not depend on a specific simulation environment; as long as the state, action, and reward functions are consistent, training can occur in any simulation setup. The training was conducted on an RTX 3090 GPU and typically completed in 15 minutes.

\subsection{Modeling Driving Maneuvers}
\label{fuzzer}
\tool\ can work with the given scenario, allowing the scenario generator to control most static factors, such as weather, map, and starting and end points for both surrounding and ego vehicles. %However, the scenario generator is also required to define the destination for the ego vehicle.

To ensure the generated safety violation cases are as realistic as possible, we defined several driving behaviors: \textit{accelerate, decelerate, brake, left lane change, and right lane change}. We used pre-defined fuzzing logic to map $v^t_{si}$ to actual driving behavior $a^t_i$ at $t$-th time step. This helps translate the MARL agent’s continuous action space into discrete driving behaviors, producing natural driving behavior. The mapping of these driving behaviors is illustrated below:
 %\jz{you need to give some reasoning why you have such mapping, give an example  for lane keeping and accelerate as decelerate is straightforward} 
%The fuzzing set representation of these driving behaviors can be denoted as:

\begin{equation}
v^t_{s_i,y} \in (0.02, 0.1] \quad \mapsto \quad a^t_i = \text{Accelerate}
\end{equation}

\begin{equation}
v^t_{s_i,y} \in [0, 0.02] \quad \mapsto \quad a^t_i = \text{Decelerate}
\end{equation}

\begin{equation}
v^t_{s_i,y} < 0 \quad \mapsto \quad a^t_i = \text{Brake}
\end{equation}

% \begin{equation}
% v^t_{s_i,x} \in [-0.01, 0.01] \quad \mapsto \quad a^t_i = \text{Lane Keeping}
% \end{equation}

\begin{equation}
v^t_{s_i,x} < -0.01 \quad \mapsto \quad a^t_i = \text{Left Lane Change}
\end{equation}

\begin{equation}
v^t_{s_i,x} > 0.01 \quad \mapsto \quad a^t_i = \text{Right Lane Change}
\end{equation}

% \begin{equation}
% \text{Brake:} \quad v_{iv} < 0
% \end{equation}

%Although there are only five modeled driving maneuvers, MARL makes real-time decisions on a second-by-second basis, creating a variety of action sequences. 

% We impose those restrictions on the Finite-State-Machine for driving behavior . Figure x indicated the Finite-State-Machine of road conditions restricted driving behavior.
\subsection{Rule-based Online Fuzzing}
The MARL works in tandem with the online fuzzer. Once the online fuzzer is triggered, it takes control of the surrounding vehicle. Here, we define several rule-based action patterns for online fuzzing.

\subsubsection{Rule-based Action Pattern}

\begin{figure}
    \centering
    \includegraphics[width=0.9\linewidth]{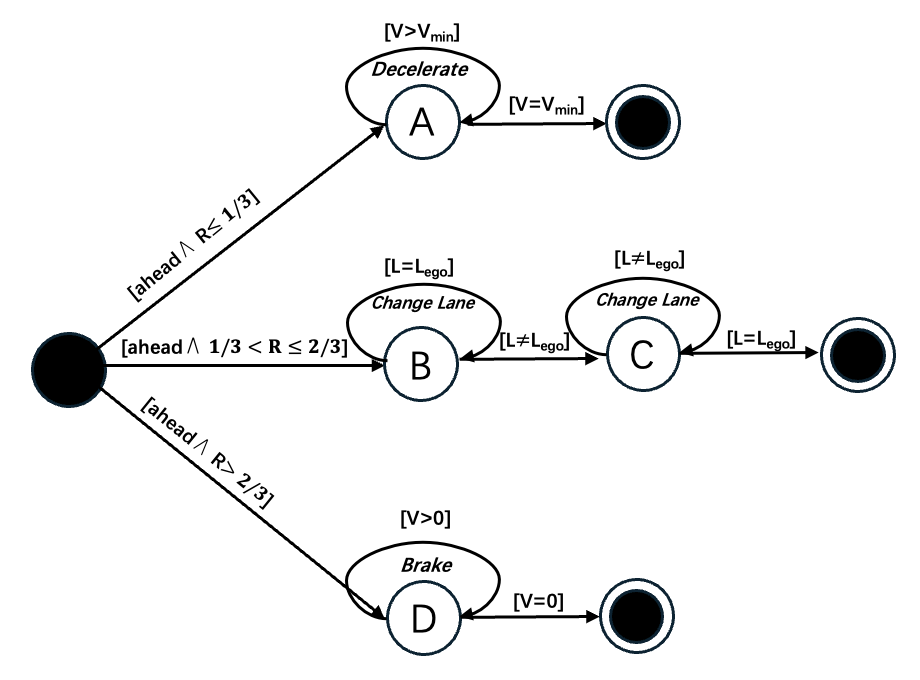}
	\caption{Ahead action pattern \cite{tian2022mosat}}
\label{ahead}  
\end{figure}

\begin{figure}
    \centering
    \includegraphics[width=0.9\linewidth]{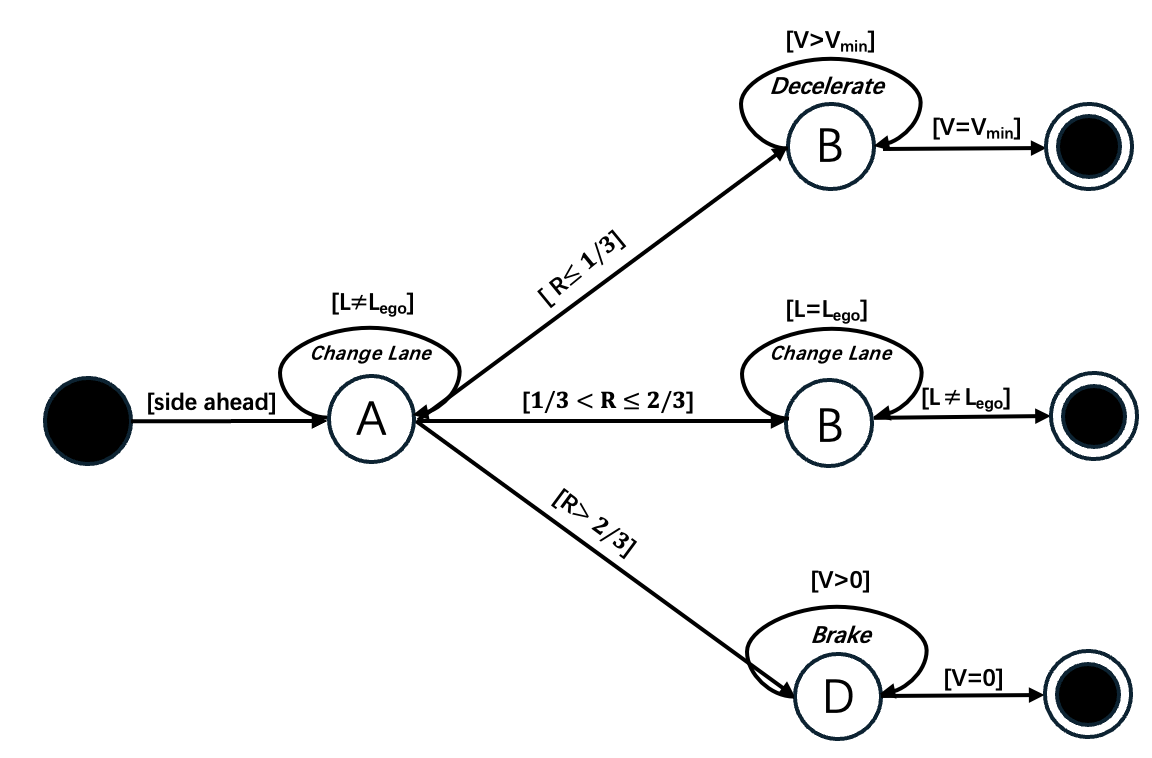}
	\caption{Side front action pattern \cite{tian2022mosat}}
\label{side front}  
\end{figure}

\begin{figure}
    \centering
    \includegraphics[width=0.9\linewidth]{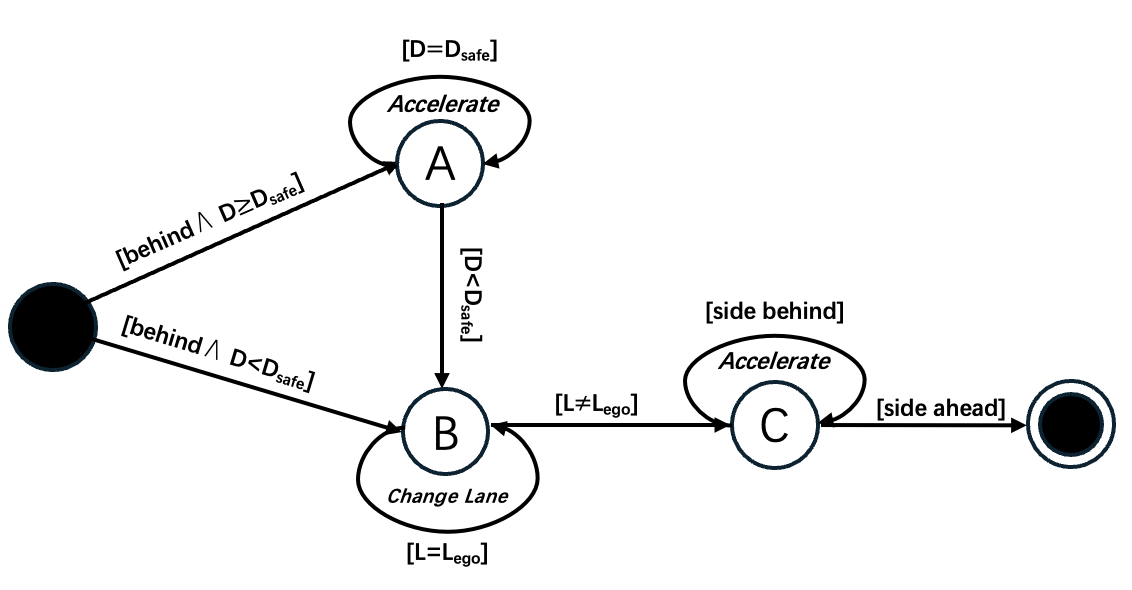}
	\caption{Behind action pattern \cite{tian2022mosat}}
\label{behind}  
\end{figure}

\begin{figure}
    \centering
    \includegraphics[width=0.9\linewidth]{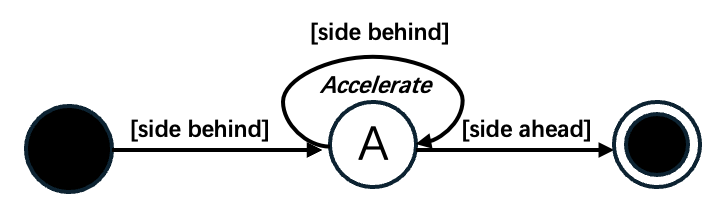}
	\caption{Side behind action pattern \cite{tian2022mosat}}
\label{side behind}  
\end{figure}

To maximize the likelihood of violations caused by the ego vehicle, we employ four rule-based action patterns, defined by \cite{tian2022mosat}, through a finite-state machine (Figures \ref{ahead} to \ref{side behind}) for the online fuzzer to manipulate. These patterns are based on previous studies of pre-crash scenarios from NHTSA (National Highway Traffic Safety Administration) \cite{nhtsa_2022_1, nhtsa_2022_2}. Once an action pattern is triggered, the surrounding vehicle will execute the behavior in the action pattern. The next $m$ time steps' driving behavior sequence will be generated at time step $t$ and can be denoted as:

\begin{equation}
    [a^t_i, a^{t+1}_i,\dots,a^{t+m}_i]
\end{equation}

The surrounding vehicle needs to finish the driving behavior in the sequence to finish the action pattern. These rule-based action patterns can be triggered in the following driving scenarios: % \jz{for these figures, if you do not draw them from scratch and copy from these cited papers, you need to put citation in the caption of each figure otherwise it is a big problem of plagiarism}:

\begin{itemize}
  \item \textbf{Ahead action pattern}: Figure \ref{ahead} shows the behavior model for the ahead action pattern, where the surrounding vehicle is in front of the ego vehicle in the same lane within the safety distance threshold $D_{\text{safe}}$ which is equal to the lane width. There are three possible driving behaviors once this action pattern is triggered: decelerate, brake, and change to the adjacent lane before returning to the original lane. %Once a driving behavior is selected, the surrounding vehicle will continue executing the behavior until it reaches the corresponding state.

    \item \textbf{Side front action pattern}: Figure \ref{side front} shows the behavior model for the side front action pattern, where the surrounding vehicle is in front of the ego vehicle but in a different lane (either left front or right front) within the safety distance threshold $D_{\text{safe}}$. The surrounding vehicle first changes to the lane where the ego vehicle is located, which is \textbf{State A}, then randomly performs one of the following maneuvers with equal probability: decelerate, change lanes, or brake.

    \item \textbf{Behind action pattern}: Figure \ref{behind} shows the behavior model for the behind action pattern, where the surrounding vehicle is behind the ego vehicle in the same lane. If the distance between the surrounding vehicle and the ego vehicle, $D$, is greater than the safety distance threshold $D_{\text{safe}}$, the finite-state machine transitions to \textbf{State A}. In this state, the surrounding vehicle will continue accelerating until $D$ becomes less than $D_{\text{safe}}$, at which point it transitions to \textbf{State B}. Then, the surrounding vehicle will change to the adjacent lane, which is \textbf{State C}, and accelerate until its relative position is in the left front or right front of the ego vehicle.

    \item \textbf{Side behind action pattern}: Figure \ref{side behind} shows the behavior model for the side behind action pattern, where the surrounding vehicle is behind the ego vehicle in a different lane (either left behind or right behind). When this pattern is triggered, the surrounding vehicle accelerates until it reaches a position in the left front or right front of the ego vehicle.

\end{itemize}

Multiple agents can trigger these patterns simultaneously. Once triggered, the online fuzzer takes control of the surrounding vehicle until the pattern is completed. The online fuzzer then decides whether to trigger the action patterns again or return control of the surrounding vehicle to the MARL.

\subsection{Driving Behavior Constraints}
\label{drc}
We impose several driving behavior constraints to ensure that surrounding vehicles can cooperate smoothly and behave as realistically as possible. Specifically, we enforce two constraints on surrounding vehicles:
\begin{itemize}
    \item \textit{Constraint 1: surrounding vehicles should not collide with other vehicles during manipulation (including the ego vehicle). }

    \item \textit{Constraint 2: surrounding vehicles should remain within road boundaries.}
\end{itemize}

These constraints are applied to both the MARL and the rule-based action pattern. Algorithm \ref{constraint} illustrates how we apply the constraint to driving behavior. In Algorithm \ref{constraint}, line \ref{input}, we first input the maximum distance between surrounding vehicle $i$ and other vehicles, $D_{\text{max}}$, the constraint threshold distance $D_{\text{constraint}}$ from empirical settings, the discrete driving behavior $a^t_i$ from the action pattern or fuzzing logic at time step $t$, and the state of surrounding vehicle $i$ at time step $t$. Then, if $D_{\text{max}}$ is smaller than $D_{\text{constraint}}$ or $S^t_i$ is outside the road boundary (line \ref{contr}), the surrounding vehicle will execute \textit{brake} at $t$-th time step (line \ref{br}), or it will output the original $a^t_i$ (line \ref{or}).

% Surrounding vehicles trigger these constraints when they come within the safety distance threshold. Once triggered, the surrounding vehicle is forced to brake. However, due to the existing motion, there is no guarantee that the surrounding vehicle will stop completely or avoid violating these constraints. In the real world, most car accidents occur due to the shared responsibility of multiple parties, which is the rationale behind this setting. 

\begin{algorithm}
\caption{Driving Behavior Constraints}
\begin{algorithmic}[1]
\label{constraint}
\STATE \textbf{Input:} $D_{max}$ (maximum distance between other vehicles), $D_{constraint}$ (constraint threshold distance), $road\_boundary$ (road boundary limits), $a^t_i$ (discrete driving behavior), $S^t_i$
\label{input}
\STATE \textbf{Output:} Updated discrete driving behavior $a^{`t}_i$
\IF{$D_{max} < D_{constraint}$ \textbf{ or } $S^t_i$ \textbf{ is outside road\_boundary}}
\label{contr}
    \STATE $a^{`t}_i = brake$
    \label{br}
\ELSE
   \STATE $a^{`t}_i = a^t_i$
   \label{or}
\ENDIF
\end{algorithmic}
\end{algorithm}

However, due to the existing motion, there is no guarantee that the surrounding vehicle will stop completely or avoid violating these constraints. This aligns with real-world situations where, even when a driver attempts to brake, it still takes time for the vehicle to come to a complete stop. This is the rationale behind this setting.

\section{Experiment}
\label{exp}
\subsection{Research Questions}
The following research questions (RQs) were assessed to evaluate the performance of \tool: 
\begin{itemize}
    % \item RQ1: How is the quality of the \tool \ dataset?
    % \item RQ2: How effectively does the \tool \ work with different multi-modal LLMs?
    \item RQ1: How effectively does the \tool \ work across different scenario and ADSs?
    
    \item RQ2: How effective is \tool \ in finding ADS safety violations compared to state-of-the-art techniques?

    \item RQ3: How realistic are the identified safety violations?
   %^ by the \tool?
    % \item RQ4: ABLATION STUDY

\end{itemize}

\subsection{Experiment Setup}

% \subsubsection{Systems Under Test}
% In the experiment, we evaluate \tool \ on two state-of-the-art ADSs: the RL-based ADS in MetaDrive \cite{li2022metadrive} and the Carla ADS in CARLA \cite{dosovitskiy2017carla}.
% \subsection{Large Language Model Selection}
We conducted all experiments using Metadrive \cite{li2022metadrive}, a powerful simulator that models various road networks and vehicle configurations with high fidelity, including different road types, lane numbers, vehicle types, and setups. Metadrive supports several end-to-end ADSs, including the Intelligent Driver Model (IDM) policy \cite{treiber2000congested} and Proximal Policy Optimization (PPO) Policy \cite{schulman2017proximal}. The IDM Policy in Metadrive automatically maintains a safe distance from moving objects and avoids static obstacles, while the PPO Policy, a three-layer MLP with a tanh activation function, is well-trained in the Metadrive environment and effectively handles most driving scenarios. These end-to-end ADSs leverage raw sensing data from the Metadrive simulator, such as LiDAR and camera inputs, to make decisions.
\par In our experiments, we assume that surrounding vehicles are initially positioned randomly near the ego vehicle within the lane, and their destinations are the same. The default number of agents is set to 3; however, \tool\ does not limit the number of agents. Each experiment is allocated a test budget of 200 runs and is conducted 5 times, with the averaged results presented. Table \ref{tab:hyperparameters} indicates the detailed settings of hyperparameter in Section \ref{method}.

\begin{table}[!ht]
\centering
\caption{Hyperparameter Settings}
\label{tab:hyperparameters}
\scalebox{0.9}{
\begin{tabular}{lll}
\hline
\textbf{Component} & \textbf{Parameter} & \textbf{Value} \\
\hline
MARL training & Neural network layers & 2 \\
& Neurons per layer & 128 \\
& Learning rate & 0.001 \\
& Initial noise & 0.75 \\
& Noise decay & 0.999995 \\
& Minimum noise & 0.01 \\
& Collision reward ($\mathbf{I}$) & 10 \\
&$d_{enclosure}$ & 0.3 \\
&$\mu_1$ & 0.7 \\
&$\mu_2$ & 0.01 \\
&$\mu_1$ & 0.5 \\
% & Training duration & 4 hours on RTX3090 GPU \\
\hline
Simulation & Lane width & 3.5 \\
& $D_{safe}$ & 3.5 \\
& $D_{constraint}$ & 2 \\
& $V_{min}$ & 2  \\

\hline
\multicolumn{3}{l}{\textit{}} \\
\end{tabular}}
\end{table}
% can be found in our supplementary material \footnote{Our supplementary material: \url{xxx}}\jz{if you still have spaces in the main paper, putting here}.

% build \tool \ on LLAVA-34B \cite{liu2024llavanext} by default, we conduct the experiment in Metadrive \cite{li2022metadrive} with the RL-based ADS. The training dataset for the LLM is encoded from the violation cases produced by Target \cite{deng2023target}. 

\subsection{Experiment Design}
\subsubsection{RQ1} 

\begin{figure}[htbp]
    \centering
    % 第一排图片
    \begin{subfigure}[b]{0.15\textwidth}
        \centering
        \includegraphics[width=\textwidth]{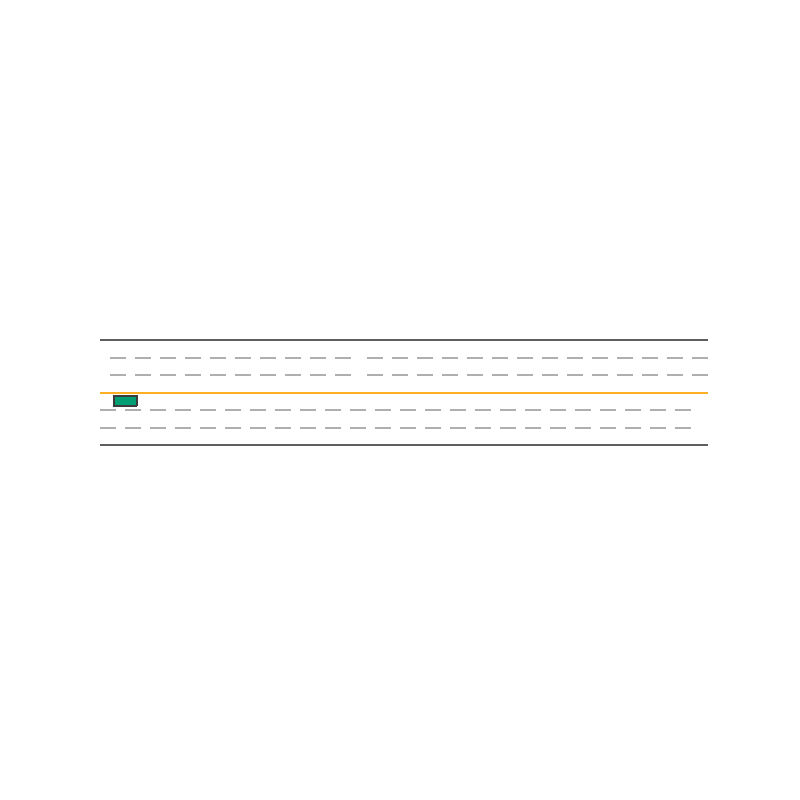} % 替换为图片路径
        \caption{Straight}
    \end{subfigure}
    \hfill
    \begin{subfigure}[b]{0.15\textwidth}
        \centering
        \includegraphics[width=\textwidth]{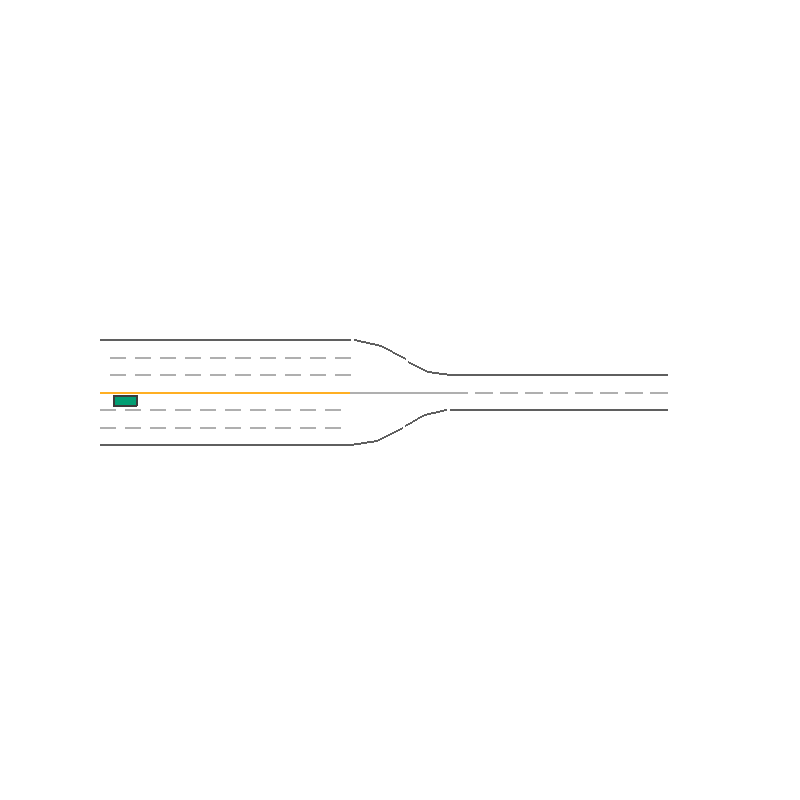} % 替换为图片路径
        \caption{Merge}
    \end{subfigure}
    \hfill
    \begin{subfigure}[b]{0.15\textwidth}
        \centering
        \includegraphics[width=\textwidth]{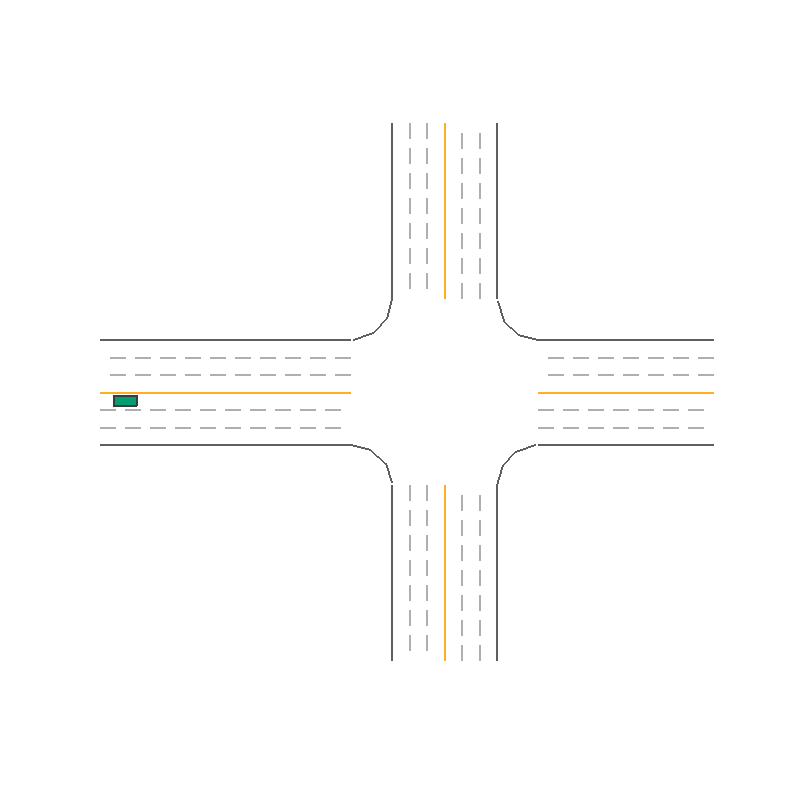} % 替换为图片路径
        \caption{Intersection}
    \end{subfigure}
    
    % \vspace{0.5cm} % 调整行间距

    % 第二排图片
    \begin{subfigure}[b]{0.15\textwidth}
        \centering
        \includegraphics[width=\textwidth]{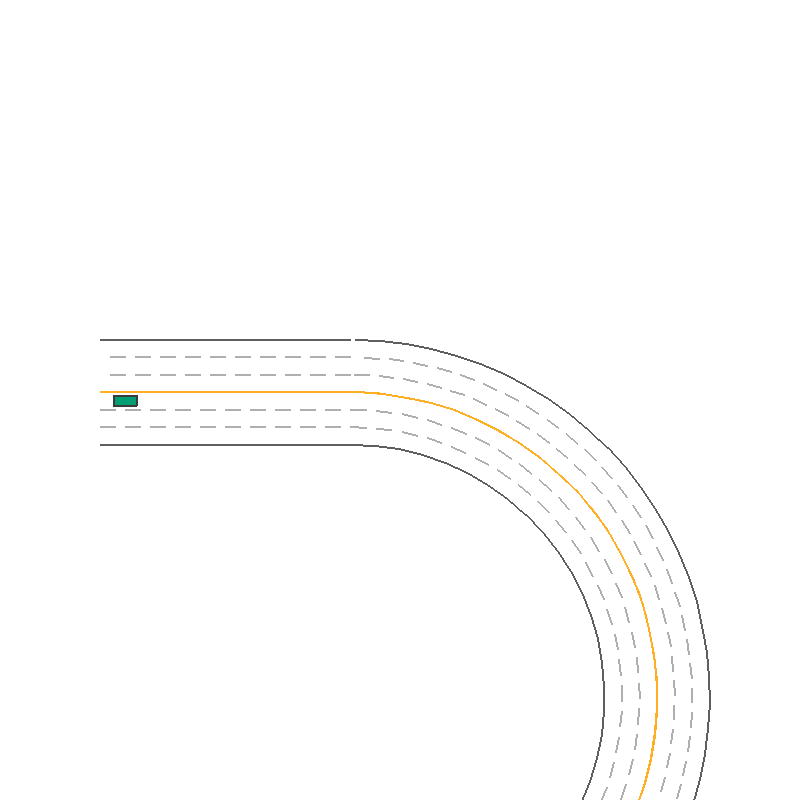} % 替换为图片路径
        \caption{Circular}
    \end{subfigure}
    \hfill
    \begin{subfigure}[b]{0.15\textwidth}
        \centering
        \includegraphics[width=\textwidth]{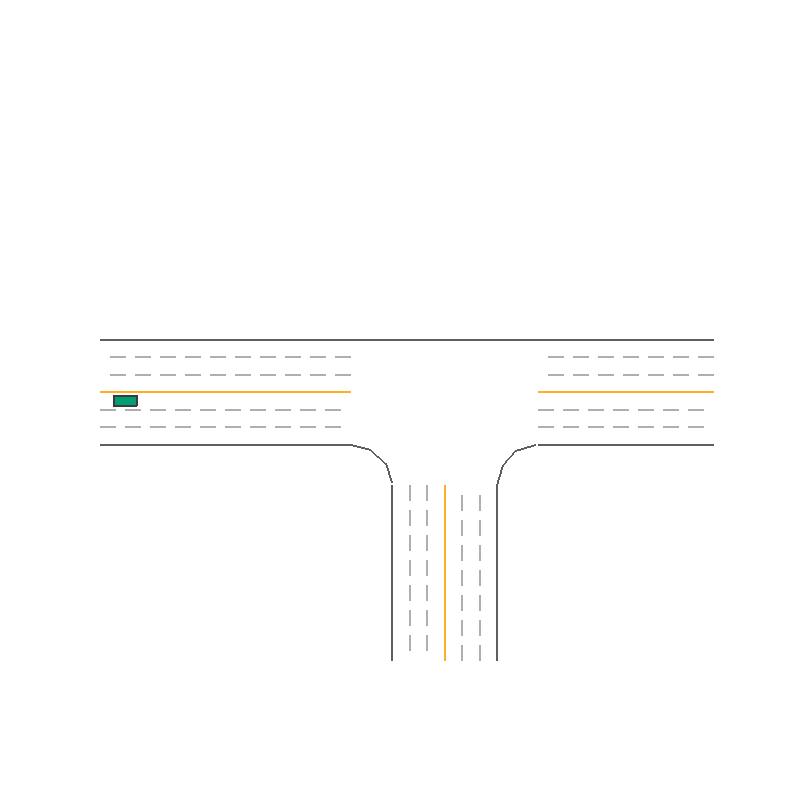} % 替换为图片路径
        \caption{T-Intersection}
    \end{subfigure}
    \hfill
    \begin{subfigure}[b]{0.15\textwidth}
        \centering
        \includegraphics[width=\textwidth]{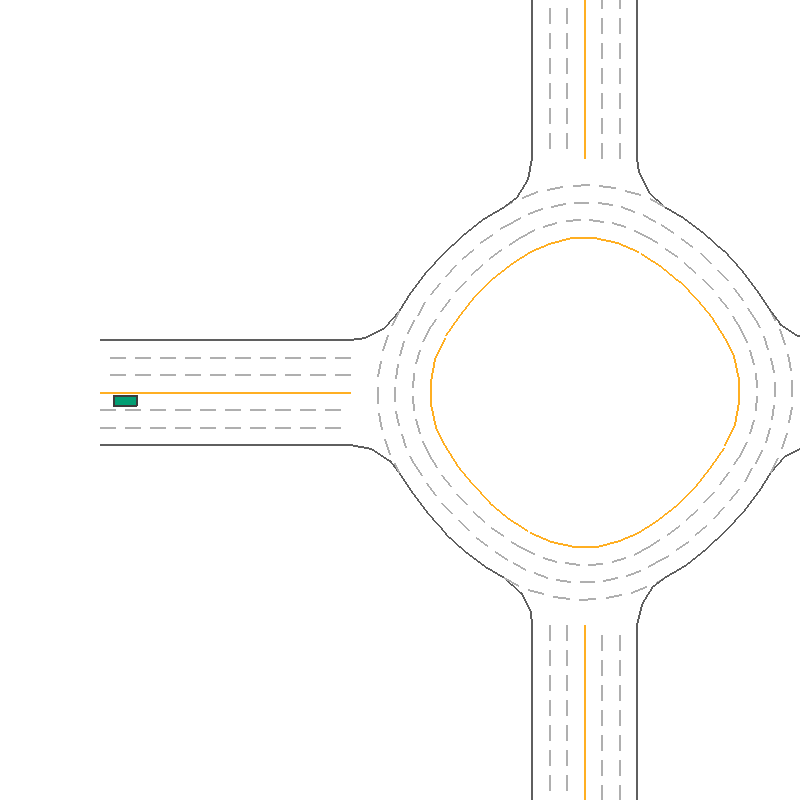} % 替换为图片路径
        \caption{Roundabout}
    \end{subfigure}

    \caption{Example of different road types}
    \label{Rexample}
\end{figure}
To evaluate the utility of \tool, we conduct a comprehensive qualitative analysis in Metadrive. First, we perform experiments across all common single road types. Figure \ref{Rexample} illustrates the road types we include, such as \textbf{Straight}, \textbf{Roundabout}, \textbf{Merge}, \textbf{T-Intersection}, \textbf{Circular}, and \textbf{Intersection}. The number of lanes is also a critical factor, as fewer lanes can lead to more crowded traffic, significantly increasing the likelihood of traffic accidents. Therefore, we test each road type with configurations ranging from 2 to 4 lanes.

Beyond single road types, Metadrive supports combinations of these blocks to simulate more complex real-world scenarios. We test combinations of randomly selected sets of 3 blocks, with lane configurations ranging from 2 to 4 lanes, to form the \textbf{MIX} scenario. \textbf{MIX} has a significantly longer length than other scenarios. All scenarios are tested with two ADSs, including the IDM Policy \cite{treiber2000congested} and PPO Policy \cite{schulman2017proximal} supported by Metadrive \cite{metadrive2024}.

The metric used to assess the utility of \tool \ is \textbf{Safety Violation Rate}. Based on \cite{liang2023rlaga}, this metric measures the percentage of safety violations detected from the ego vehicle during the test. Here we define \textbf{safety violation} \cite{deng2023target}:
\begin{itemize} 
\item \textbf{Multiple vehicles crash}: A crash involving the ego vehicle caused by the complex interplay among at least two surrounding vehicles within $D_{constraint}$.

\item \textbf{Abnormal ego vehicle trajectory}: The trajectory of the ego vehicle is extremely abnormal caused by at least two surrounding vehicles within $D_{constraint}$, posing a danger on the road and likely to cause an accident.

\end{itemize} 

% which may include crashes involving multiple vehicles or abnormal trajectories such as reverse driving \cite{deng2023target}.

% \begin{itemize} 
% \item \textbf{Unsafe Behavior Rate}: 
% Based on \cite{liang2023rlaga}, this metric measures the total number of unsafe behaviors detected from the ego vehicle during the test, which may include crashes involving multiple vehicles or abnormal trajectories such as reverse driving.

% \item \textbf{TOP-K}: Based on \cite{deng2022scenario}, this efficiency metric measures the number of test cases required to identify the first \textit{K} unsafe behavior cases, with K=10 in this paper. \end{itemize}

% We summarized the unsafe behaviour manually from the recording.

\subsubsection{RQ2}
In RQ2, we compared \tool \ with three state-of-the-art (SOTA) baselines in Metadrive: \textit{Random}, \textit{GA}\cite{tian2022mosat,li2020av}, and \textit{single-agent RL \cite{liang2023rlaga}}. To ensure a fair comparison, we set the same number of surrounding vehicles across all baselines, allowing each approach to control these surrounding vehicles. The detailed settings for each baseline are listed below:
\begin{enumerate}

    \item \textit{Random}: In this baseline, the actions of each surrounding vehicle at each time step are randomly selected from the defined driving maneuvers in Section \ref{fuzzer}.
    
    \item \textit{GA offline fuzzer}  \cite{tian2022mosat,li2020av}: This baseline leverages a predefined action sequence offline. The action sequences are randomly initialized at the beginning and controlled by NSGA-II \cite{deb2000fast} through crossover and mutation. The fitness function of the GA is based on safety violations caused by the ego vehicle; if a safety violation occurs, the fitness value of the population is set to 20. The generation size is 10, and the number of generations is 10, aligning with a testing budget of 200.
     
    \item \textit{Single-agent RL}  \cite{liang2023rlaga}: This baseline allows multiple single-agent RL models to control multiple surrounding vehicles independently, without communication between them. The agent's model structure and training environment are the same as \tool; however, all cooperative blocks in MARL are removed to train this single-agent RL. We ensure that the agent has converged during training. The single-agent RL model uses the same rule-based driving maneuvers mapping as described in Section \ref{fuzzer}.

    %fuzzing logic as \tool.
\end{enumerate}

We compared \tool\ with baselines in all scenarios defined in RQ1 using a 4-lane road and the IDM policy ADS. The 4-lane road presents a larger state space, and the IDM policy ADS is more robust than the PPO policy ADS, creating a greater challenge for the testing technique. In RQ2, we introduce another metric, \textbf{TOP-K} \cite{deng2022scenario}. This efficiency metric measures the number of safety violation cases required to identify the first \textit{K} safety violation cases, with K=5 in this paper.
\subsubsection{RQ3}

To validate the realism of our generated violations, we conducted a user study. We randomly selected 5 safety violations generated by \tool \ from each scenario in RQ1 and invited 5 experienced ADS testers to participate. Testers were asked to evaluate the realism of the generated scenarios using a rating scale from 1 to 5, where 1 indicated `not realistic' and 5 indicated `very realistic'.

% Figure \ref{int} indicates the interface of our user study. 

% \begin{figure}
%     \centering
%     \includegraphics[width=1\linewidth]{img/user_study_int.jpg}
% 	\caption{User study interface.}
% \label{int}  
% \end{figure}
% \par Moreover, we selected several safety violations in both top-down view and 3D view to conduct a case study to justify the realism of our generated safety violations.

% In RQ3, we compare \tool \ with GA-based scenario generator in RQ1 with other SOTA testing cases generation baselines including: GA-based method MOSAT, Online-RL based method, and RL-GA based work GARL, and online fuzzer RL, and Random. We compare those methods in the metadrive simulator and the ego-vehicle is equipped with a rl-based ADS. We compare those baslines in the following aspect: violation rate in 400 runs, diversity in generated scenario and violation types, top-10,
\section{Result}
\label{result}

\subsection{Utility of \tool \ (RQ1)}
\begin{table}[ht]
\centering
\caption{
%ADSs' averaged safety violation rate (\%) across different scenarios, the number after the ADS indicates the lane number, more robust ADS is marked in bold.
Averaged safety violation rate (\%) of ADSs across different scenarios. The number following the ADS indicates the lane count (e.g., L2 denotes 2 lanes). For each road type and lane count, the lowest averaged safety violation rate is highlighted.
}
\scalebox{0.8}{
\begin{tabular}{|c|c|c|c|c|c|c|c|c|c|}
\hline
\textbf{Road Type} & \textbf{IDM L2} & \textbf{PPO L2} & \textbf{IDM L3} & \textbf{PPO L3} & \textbf{IDM L4} & \textbf{PPO L4} \\ \hline
\textbf{Straight} & \textbf{21.1} & 26.6 & \textbf{15.1} & 24.6 & \textbf{10.1} & 19.3 \\ \hline
\textbf{Roundabout} & \textbf{35} & 40.5 & \textbf{16.8} & 34.0 & \textbf{10.3} & 27.7 \\ \hline
\textbf{Merge} & \textbf{32.3} & 35.5 & \textbf{19.0} & 26.5 & \textbf{11.1} & 23.0 \\ \hline
\textbf{T-Intersection} & \textbf{27.3} & 45.7 & \textbf{12.5} & 38.3 & \textbf{11.1} & 30.8 \\ \hline
\textbf{Circular} & \textbf{29.3} & 32.5 & \textbf{16.5} & 30.0 & \textbf{14.9} & 23.3 \\ \hline
\textbf{Intersection} & \textbf{28.5} & 43.7 & \textbf{14.0} & 41.5 & \textbf{9.3} & 26.5 \\ \hline
\textbf{MIX} & \textbf{24.5} &32.5 & \textbf{18.5} & 34.5 & \textbf{13.0} & 19.0 \\ \hline
\end{tabular}
}
 
\label{RQ1L2}
\end{table}

Table \ref{RQ1L2} shows the utility results of \tool\ on 2- to 4-lane roads, demonstrating its effectiveness in identifying safety violations across all scenarios and ADS models. Notably, as the number of lanes increases and the road becomes less crowded, the safety violation rate decreases, which aligns with our intuition.

\par Then, we find that the IDM policy ADS performs better than the PPO policy ADS in all scenarios, which aligns with the official documentation of Metadrive. The IDM policy incorporates a rule-based approach to maintain distance from moving objects and automatically sidestep static objects, allowing it to make more robust decisions when facing high challenges from \tool. In contrast, the PPO policy is RL-based, and during training, it may not encounter such challenging situations, resulting in lower robustness when tested with \tool.

% \par Moreover, looking into different road types, we found \textbf{Intersection} has the least safety violation rate for IDM policy in 4-lane road, which could the most challenge scenario for IDM poliy this is because xx\jz{missing interpretation}.

\begin{figure}[ht]
\centering

\begin{subfigure}[b]{0.5\textwidth}
\centering
\includegraphics[width=0.32\linewidth]{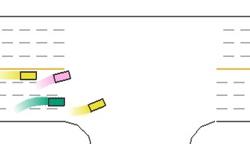}
\includegraphics[width=0.32\linewidth]{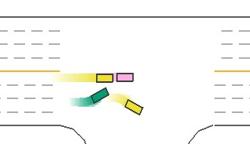}
\includegraphics[width=0.32\linewidth]{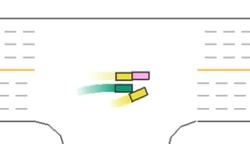}
\caption{Example for multiple vehicles crash I}
\label{c1}
\end{subfigure}

\begin{subfigure}[b]{0.5\textwidth}
\centering

\includegraphics[width=0.32\linewidth]{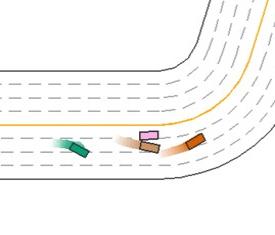}
\includegraphics[width=0.32\linewidth]{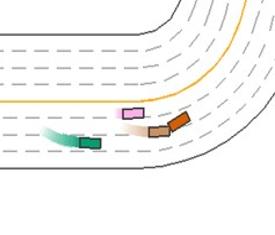}
\includegraphics[width=0.32\linewidth]{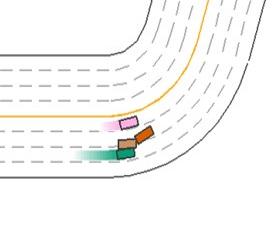}
\caption{Example for multiple vehicles crash II}
\label{c2}
\end{subfigure}

% \begin{subfigure}[b]{0.5\textwidth}
% \centering

% \includegraphics[width=0.32\linewidth]{img/crash/crash_image_10_resized.jpg}
% \includegraphics[width=0.32\linewidth]{img/crash/crash_image_11_resized.jpg}
% \includegraphics[width=0.32\linewidth]{img/crash/crash_image_12_resized.jpg}
% \caption{Example for multiple vehicles crash III}
% \label{c3}
% \end{subfigure}

% \begin{subfigure}[b]{0.5\textwidth}
% \centering

% \includegraphics[width=0.45\linewidth]{img/crash/backhit_image_1_resized.jpg}
% % \includegraphics[width=0.32\linewidth]{img/crash/backhit_image_2_resized.jpg}
% \includegraphics[width=0.45\linewidth]{img/crash/backhit_image_3_resized.jpg}
% \caption{Example for multiple vehicles crash IV}
% \label{c4}
% \end{subfigure}

% \caption{Examples of violation from simulation (Part 1)}

% \begin{subfigure}[b]{0.5\textwidth}
% \centering
% \includegraphics[width=0.49\linewidth]{img/reverse/reverse_1.jpg}
% \includegraphics[width=0.49\linewidth]{img/reverse/reverse_2.jpg}
% \caption{Example for abnormal ego vehicle trajectory I}
% \label{r1}
% \end{subfigure}

\begin{subfigure}[b]{0.5\textwidth}
\centering

\includegraphics[width=0.45\linewidth]{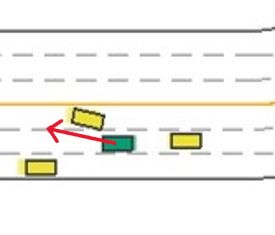}
\raisebox{0.5mm}{\includegraphics[width=0.45\linewidth]{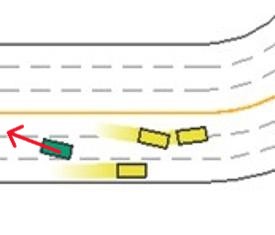}}
\caption{Example for abnormal ego vehicle trajectory I}
\label{r2}
\end{subfigure}

\begin{subfigure}[b]{0.5\textwidth}
\centering
\includegraphics[width=0.3\linewidth]{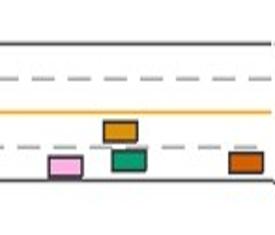}
\includegraphics[width=0.3\linewidth]{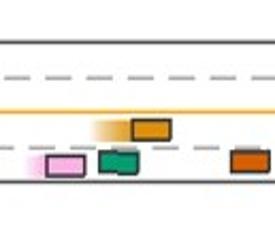}
\raisebox{1.2mm}{\includegraphics[width=0.29\linewidth]{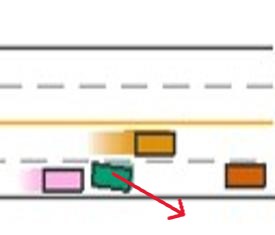}}  % 将第三张图片上移5mm
\caption{Example for abnormal ego vehicle trajectory II}
\label{r3}
\end{subfigure}

\caption{Examples of safety violation from top-down view simulation}
\label{Examplessim2}
\end{figure}

\begin{figure}[ht]
\centering

\begin{subfigure}[b]{0.49\textwidth}
\centering

\includegraphics[width=0.45\linewidth]{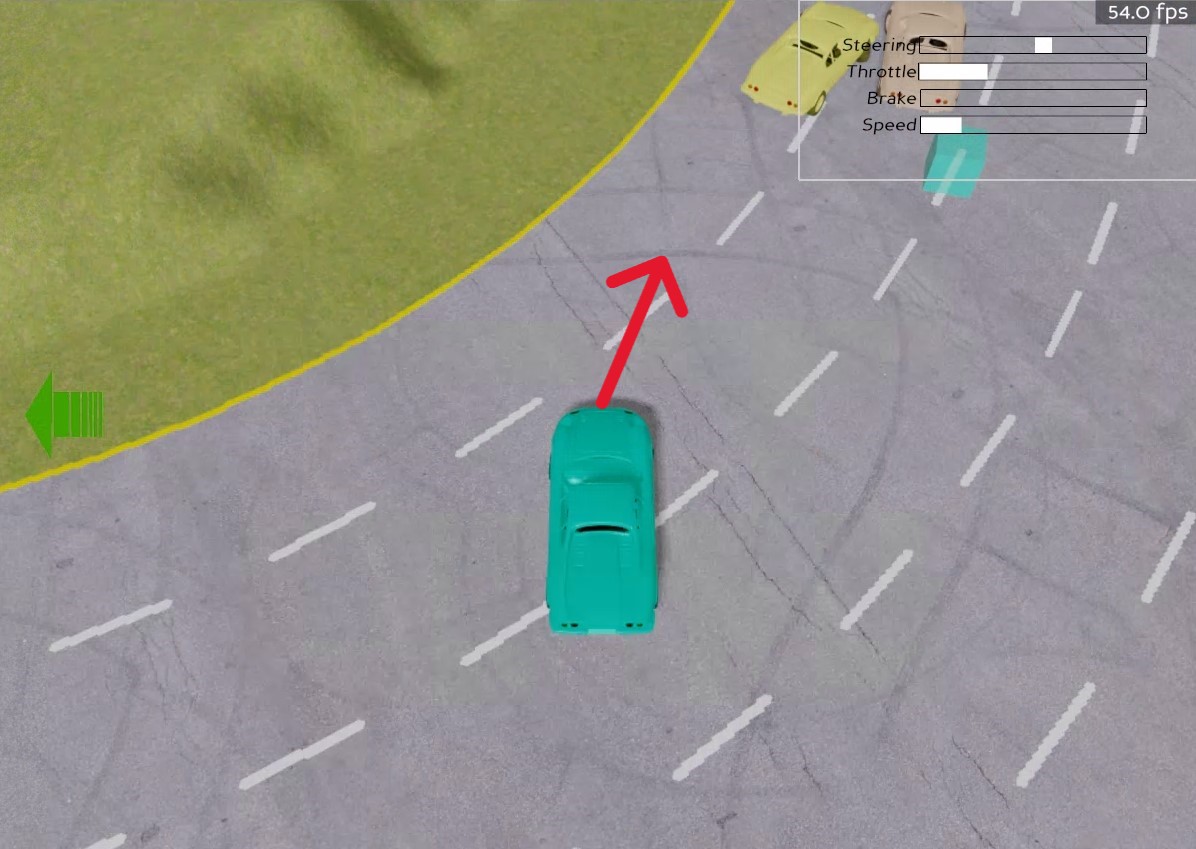}
\includegraphics[width=0.45\linewidth]{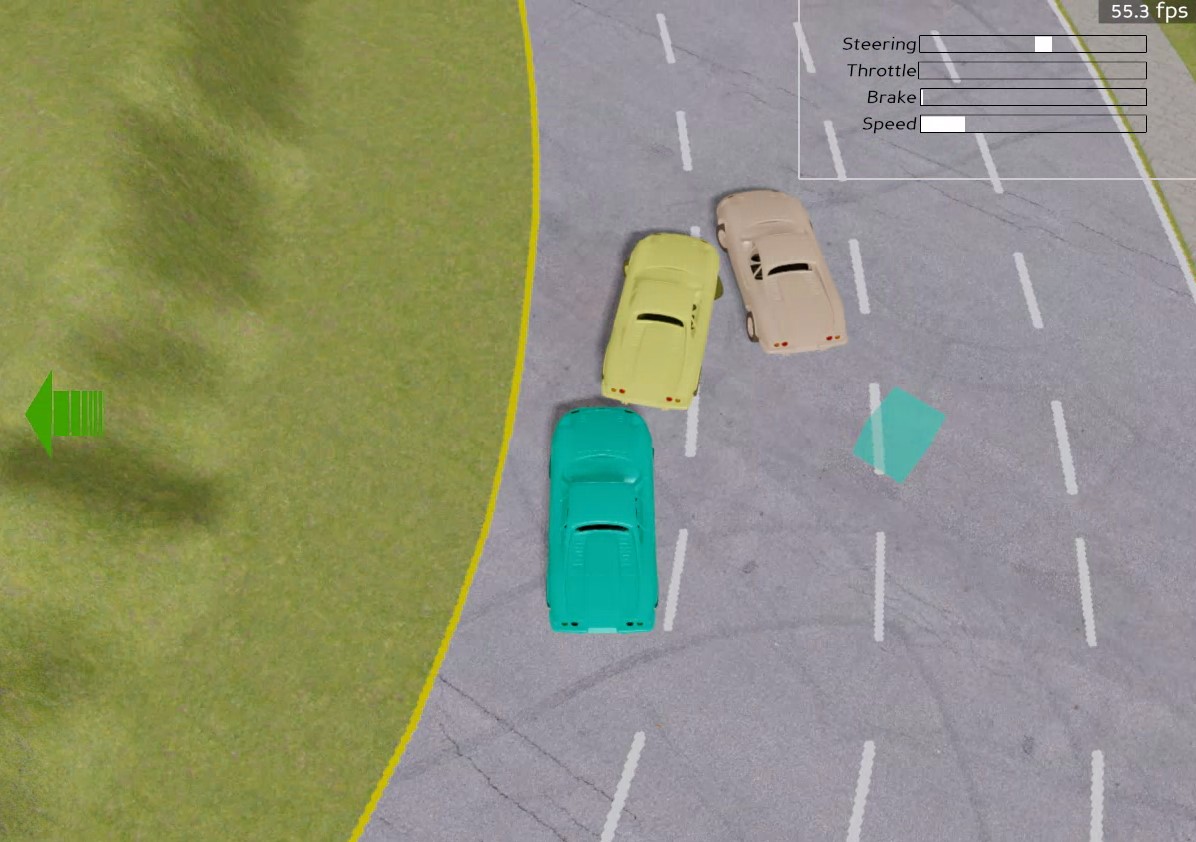}
\caption{Example for multiple vehicles crash in 3D view}
\label{3dr2}
\end{subfigure}

\begin{subfigure}[b]{0.49\textwidth}
\centering
\includegraphics[width=0.45\linewidth]{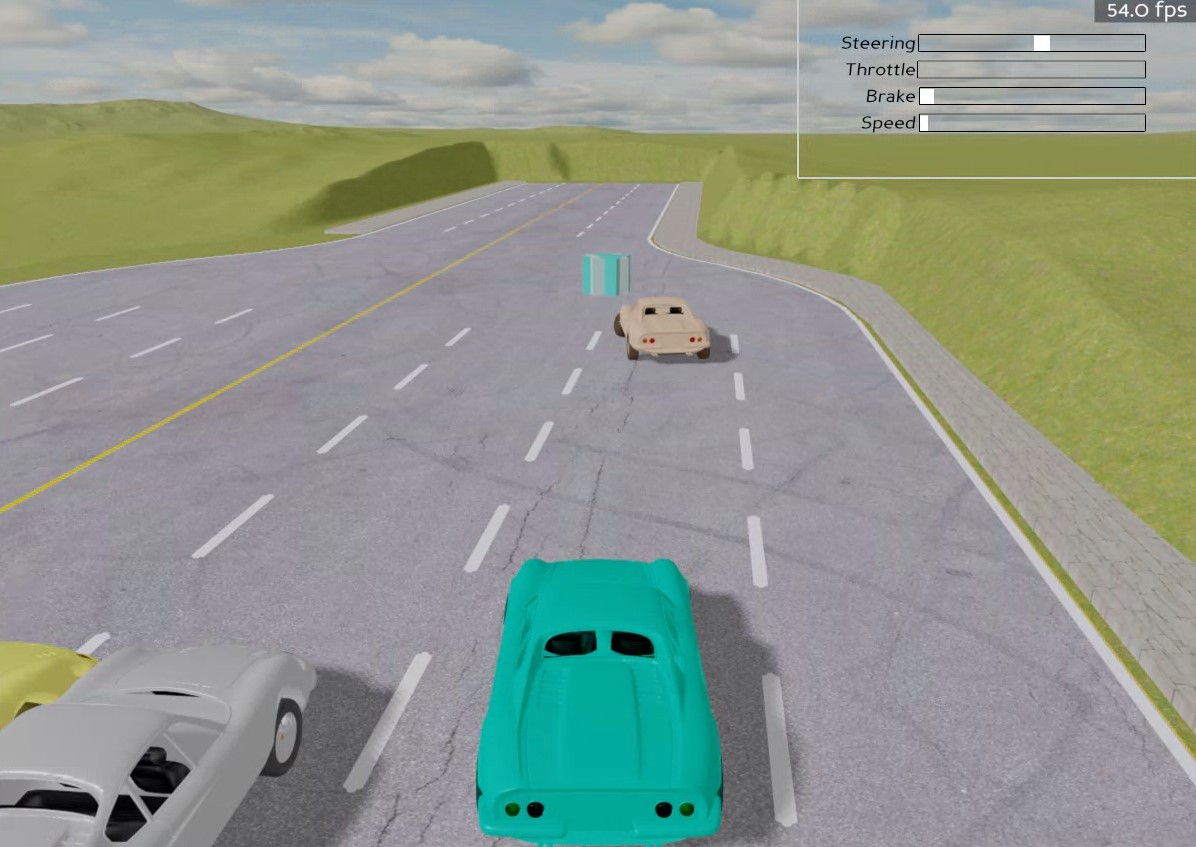}
\includegraphics[width=0.45\linewidth]{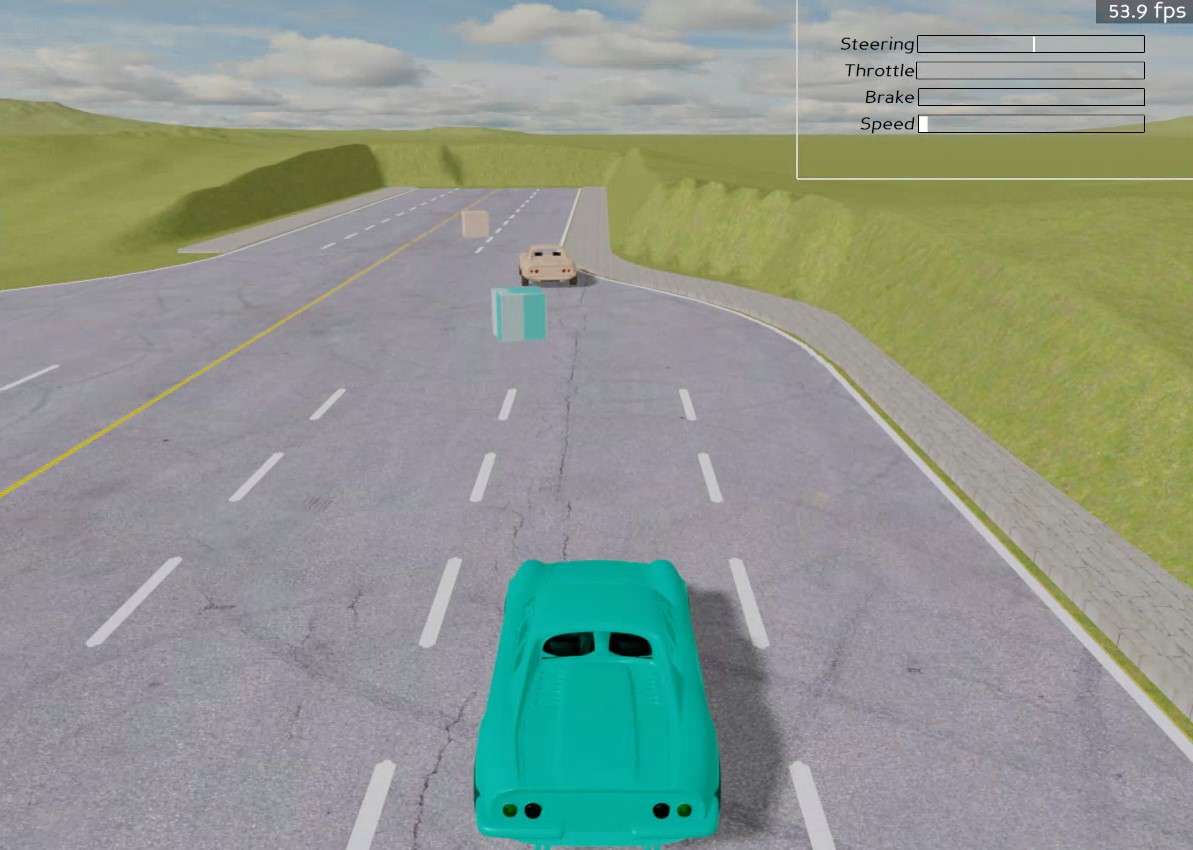}

\caption{Example for abnormal ego vehicle trajectory in 3D view}
\label{3dc3}
\end{subfigure}

\begin{subfigure}[b]{0.49\textwidth}
\centering
\includegraphics[width=0.45\linewidth]{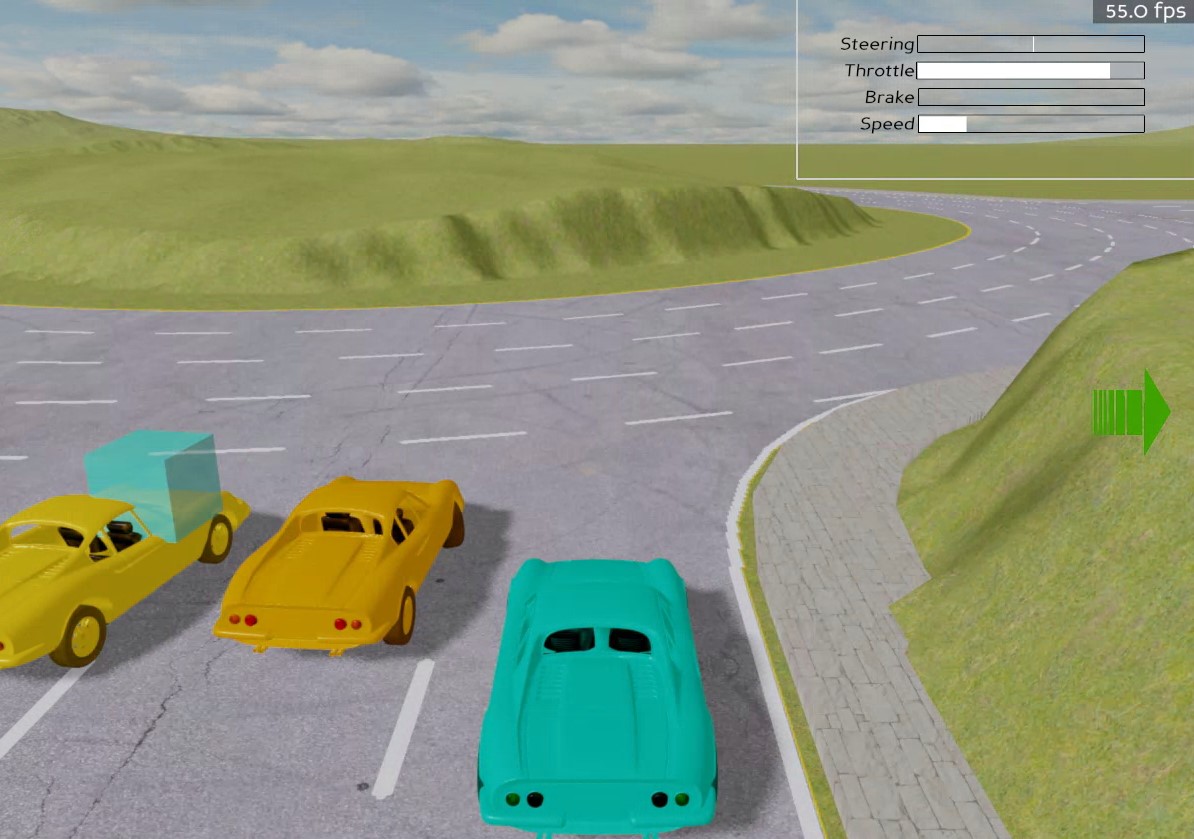}
\includegraphics[width=0.45\linewidth]{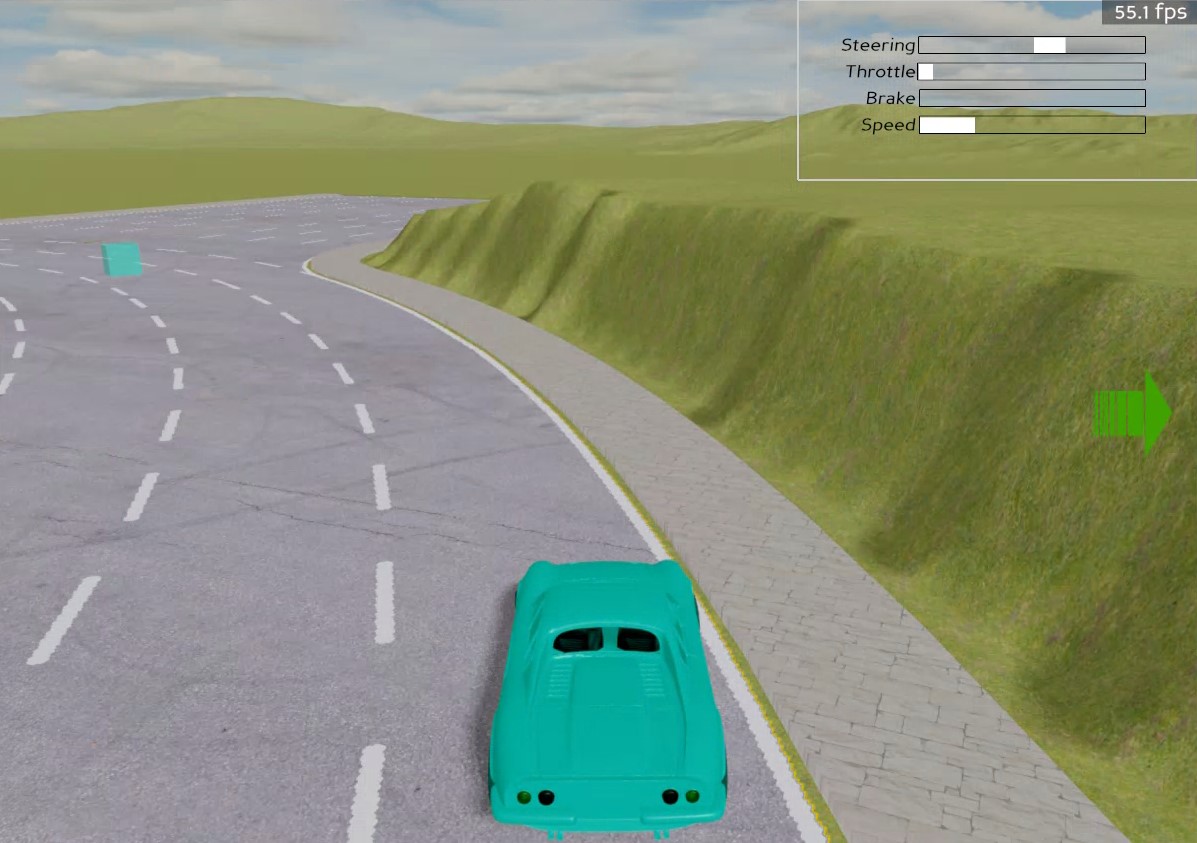}

\caption{Example for abnormal ego vehicle trajectory in 3D view II}
\label{3dc4}
\end{subfigure}

\caption{Examples of safety violation from 3D view simulation}
\label{Examplessim3}
\end{figure}

We manually analyze the collected records and select some typical examples for a case study. Figure \ref{Examplessim2} presents examples of safety violations from the simulation using a top-down view, and Figure \ref{Examplessim3} presents two more examples using a 3D view, with each example elaborated on below as a case study.
%shows examples of safety violations from the simulation, using a top-down view.

% \begin{itemize} 
% \item \textbf{Multiple vehicles crash}: A crash involving the ego vehicle caused by cooperation or chain reactions between at least two NPC vehicles. 
% \item \textbf{Abnormal ego vehicle trajectory}: The trajectory of the ego vehicle is extremely abnormal, posing a danger on the road and likely to cause an accident.

% \end{itemize} 

\subsubsection{Case Study I}
Figure \ref{c1} depicts a multiple-vehicle crash in the \textbf{Intersection} scenario. The green block represents the ego vehicle, while other blocks represent surrounding vehicles. Initially, the ego vehicle is accelerating, and the front yellow vehicle attempts to change into its lane. The ego vehicle tries to avoid the surrounding vehicle by switching to the left lane, but the yellow vehicle again attempts to change into the same lane. However, two other surrounding vehicles are positioned on the ego vehicle's left side, leaving no space for the ego vehicle to change lanes, resulting in a collision with the yellow lane-changing vehicle.

\subsubsection{Case Study II}
Figure \ref{c2} illustrates the \textbf{MIX} scenario, where a pink surrounding vehicle is stopped in the lane, and a brown surrounding vehicle is driving on the same lane. As the brown vehicle approaches the pink vehicle, it changes to the adjacent lane to continue moving forward. Unfortunately, after changing lanes, the brown vehicle collides with a red vehicle and comes to a stop. To avoid the two collided vehicles, the green ego vehicle attempts to change lanes; however, it is too late, and the ego vehicle collides with the stopped brown vehicle.

\subsubsection{Case Study III}

In Figure \ref{r2}, the yellow surrounding vehicle to the left of the green ego vehicle is attempting to change into the same lane. Additionally, another yellow surrounding vehicle is positioned ahead of the ego vehicle. The ego vehicle makes an abnormal and dangerous decision to reverse. If there were a vehicle behind the ego vehicle, this could likely result in a collision. In some countries, reversing on the road is strictly prohibited \cite{npc_law}.

\subsubsection{Case study IV}

In Figure \ref{r3}, a brown surrounding vehicle is passing the green ego vehicle quickly, while a pink surrounding vehicle behind the ego vehicle is accelerating and approaching it. A red surrounding vehicle is stopped ahead of the ego vehicle. In this situation, the ego vehicle makes an abnormal decision to drive directly off the road to avoid collision.

%Figure \ref{Examplessim3} indicates two safety violation examples from 3D view simulation.
\subsubsection{Case Study V}
In Figure \ref{3dr2}, two surrounding vehicles are crashed in front of the green ego vehicle. Despite the collision, the ego vehicle maintains its course and attempts to pass the yellow surrounding vehicle, believing there is enough space on the left. Unfortunately, this leads to another collision involving multiple vehicles.

\subsubsection{Case Study VI}
Figure \ref{3dc3} indicates a \textbf{Merge} scenario, two surrounding vehicles collided beside the green ego vehicle, and a brown surrounding vehicle is moving forward in front of the ego vehicle. However, even though the brown vehicle has already moved far away, the ego vehicle still does not move, which causes a timeout issue.

\subsubsection{Case Study VII}

Figure \ref{3dc4} indicates a \textbf{Roundabout} scenario, initially, two yellow surrounding vehicles are moving on the left-hand side of the green ego vehicle, which is in the outermost lane of the road. They are preparing to enter the roundabout by slightly turning right. However, due to the close proximity of the surrounding vehicles, the ego vehicle oversteers and drives off the road.
%\jz{it would be even better to show the recorded video and logs to the developers of these driving models, probably UCLA to confirm whether it is a known bug or help them to find bug. System reviewers would love it}

\par 
From the case study, we observe that the ADSs under test can generally respond to the dynamic trajectory of a single surrounding vehicle in most of cases. However, in these scenarios with complex dynamic interactions among multiple surrounding vehicles, even when the ego vehicle responds promptly, it is still unable to avoid safety violations. We have identified these scenarios as potentially containing bugs in the autonomous driving model and have shared the corresponding videos and logs with the developers of MetaDrive to support system improvement. More safety violation demos can be found in our anonymous repository \footnote{https://anonymous.4open.science/r/MARL-OT-E1B5}.
%\tochecklf{From the case study, we observe that the ADS responds to the incoming manipulation of surrounding vehicles in all cases, highlighting the robustness of the system. However, with the cooperation among surrounding vehicles, even the ego vehicle has corresponding responses, it is still unable to avoid safety violations. We have selected cases that may contain bugs in the system and shared the videos and logs with the developers of Metadrive to assist in improving the system.}
% \jz{this is good, you can walk through these scenarios, but I would not call them realism, instead I will put them in the RQ1 to show what kind of safety violations we can detected. In RQ3, you shall show the boxplot of the user agreement of the realism of the generated scenarios. In our TSE paper, we use 1-7 scales, you can use the same and show the boxplot which shows the median, best and worst case. To save time, you can run them with 3 raters, you can call experienced ADS testers}

\subsection{Comparison to Baselines (RQ2)}
Table \ref{RQ2} shows the performance of \tool \ and baseline methods in the IDM policy ADS across different scenarios. As indicated in RQ1, we selected the IDM policy for testing in RQ2 due to its demonstrated robustness. We observe that \tool \ outperforms all other baseline methods across all metrics, with a specific improvement of up to 136.2\% over Single-Agent RL. The GA and Random baselines identify very few safety violations in some scenarios.

Different methods show significant performance differences across various maps. In the \textbf{Merge} scenario, even the Random baseline achieves a 5.4\% safety violation rate, as this scenario involves multiple vehicles entering a narrower road, increasing the likelihood of safety violations. In other words, the \textbf{Merge} scenario is easier to trigger safety violations and more dangerous. In this scenario, we observe that both the Single-Agent RL and \tool \ nearly match each other's performance. GA also shows notable improvement, due to its evolutionary nature, which carries forward safety violations found in the initial generation throughout the process.

\par In less dangerous scenarios such as \textbf{Straight}, \textbf{Circular}, and \textbf{MIX}, the safety violation rate of the GA and Random baselines is under 1\%, while RL-based online methods remain consistently stable. This reveals a disadvantage of GA: it tends to retain the chromosome representations that lead to safety violations. If no safety violations are found during the initial generation's random initialization, GA relies on random mutation and crossover, which cannot guarantee that evolution will approach the violation boundary. As a result, GA struggles to adapt quickly to scenarios. In contrast, online methods like RL can better adapt to the scenario and identify violations, highlighting the limitations of offline methods like GA in online testing scenarios.

\par When we look at the online testing method RL, although it remains stable in generating violations across different scenarios, without agent cooperation, single-agent RL still has a large gap with \tool\ when generating safety violations involving multiple vehicles. The results also demonstrate that offline methods like GA and Random can only identify a limited number of safety violations, underscoring the challenge of detecting subtle errors in seemingly robust ADSs like IDM. This highlights the need for our approach, which generates scenarios involving multiple surrounding vehicles with complex interactions.

%\tochecklf{\par The results also show that offline methods like GA and Random can only identify a few safety violations, highlighting the robustness of the IDM ADS and the difficulty in generating safety violations within such a system.}

\begin{table}[ht]
\centering
\caption{Quantitative result of safety violation case generating efficiency for different methods, `None' indicates TOP-5 cannot be found, best is marked in bold.}
\renewcommand{\arraystretch}{1.5} % Optional: Adjust row height
\scalebox{0.8}{
\begin{tabular}{|c|c|c|c|c|c|}
\hline
\textbf{Map Type} & \textbf{Metric} & \textbf{\tool} & \textbf{GA} & \textbf{ RL} & \textbf{Random} \\ \hline
\multirow{2}{*}{Straight} 
    & Violation Rate (\%) & \textbf{10.1} & 0.3 & 5.2 & 0.1 \\ \cline{2-6}
    & TOP-5             & \textbf{68.8} & None & 120.4 & None\\ \hline
\multirow{2}{*}{Roundabout} 
    & Violation Rate (\%) & \textbf{10.3} & 1.9 & 5.1 & 2.2 \\ \cline{2-6}
    & TOP-5             & \textbf{69.2} & 200.0 & 129.8 & 173.7 \\ \hline
\multirow{2}{*}{Merge} 
    & Violation Rate (\%) & \textbf{11.1} & 6.7 & 11.0 & 5.4 \\ \cline{2-6}
    & TOP-5            & \textbf{56.4} & 88.0 &58.2 & 114.2 \\ \hline
\multirow{2}{*}{T-Intersection} 
    & Violation Rate (\%) & \textbf{11.1} & 2.9  & 4.7 & 2.4 \\ \cline{2-6}
    & TOP-5            & \textbf{55.2} & 141.3 & 105.8 & 194.3 \\ \hline
\multirow{2}{*}{Circular} 
    & Violation Rate (\%) & \textbf{14.9} & 0.2 & 7.8 & 0.6 \\ \cline{2-6}
    & TOP-5            & \textbf{36.8}& None & 79.2 & None \\ \hline
\multirow{2}{*}{Intersection} 
    & Violation Rate (\%) & \textbf{9.3} & 1.8 & 8.2 & 1.7 \\ \cline{2-6}
    & TOP-5             & \textbf{38.6} & None & 85.6 & None \\ \hline
\multirow{2}{*}{MIX} 
    & Violation Rate (\%) & \textbf{13.0} & 0.5 & 7.9 & 0.8 \\ \cline{2-6}
    & TOP-5             & \textbf{50.0} & None & 72.6 & None \\ \hline
\end{tabular}
}

\label{RQ2}
\end{table}

\subsection{Realism of Generated Cases (RQ3)}

\begin{figure}[ht]
    \centering
    \includegraphics[width=1\linewidth]{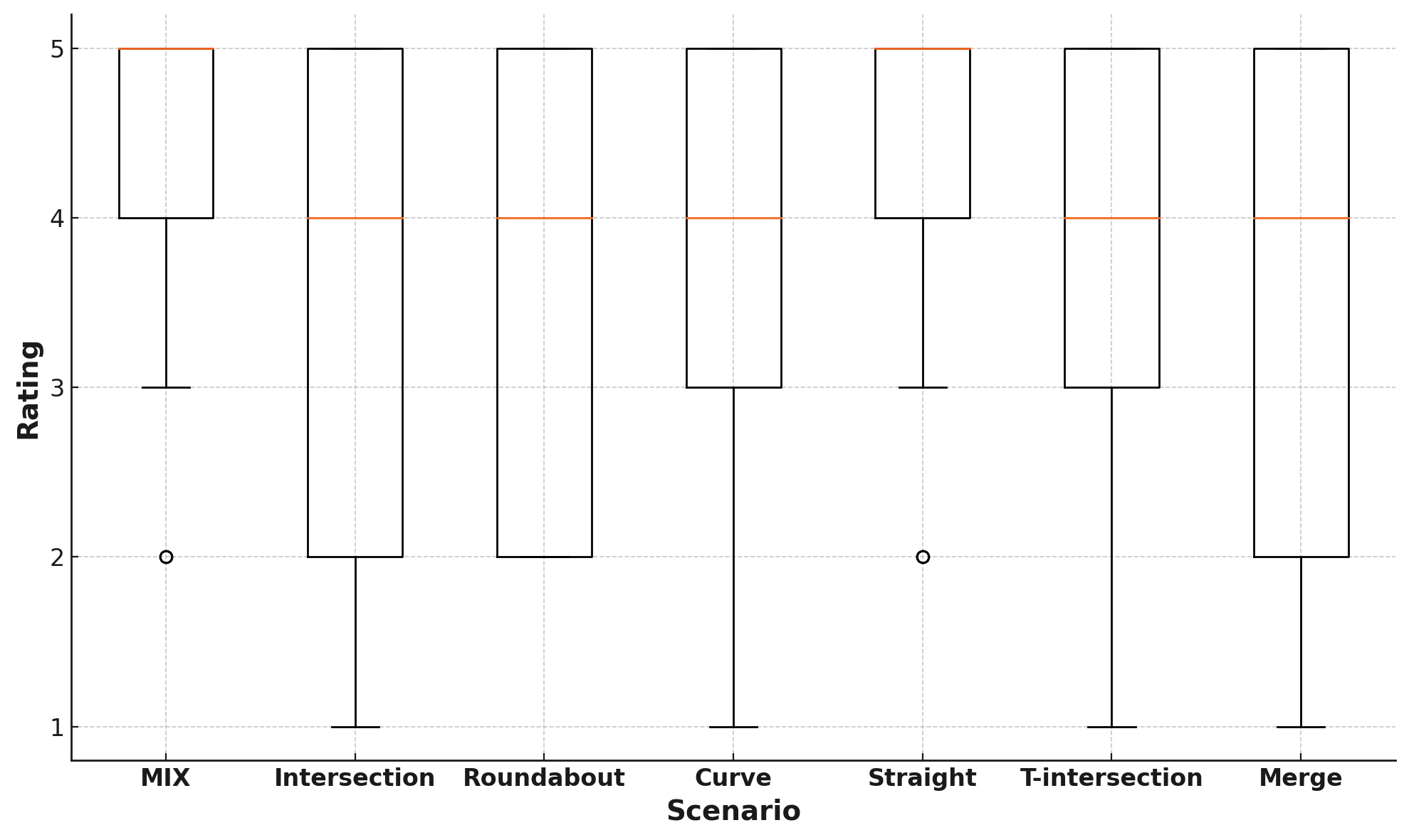}
	\caption{Boxplots for User Ratings on Different Scenarios with Outliers}
\label{user}  
\end{figure}
Figure \ref{user} shows the box plot of realism ratings from the user study, where the red line indicates the median rating in each scenario. We find that the median rating for all scenarios is above 4, indicating that our generated scenarios have high realism. However, there are still some outliers, and here we further analyze those outliers.

\subsubsection{Outlier Analysis I}
\begin{figure}[ht]
\centering

\includegraphics[width=0.32\linewidth]{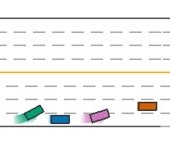}
\raisebox{1mm}{\includegraphics[width=0.32\linewidth]{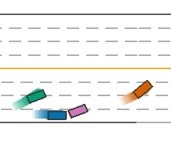}}
\raisebox{1mm}{\includegraphics[width=0.32\linewidth]{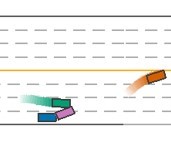}}

\caption{Outlier in the MIX scenario}
\label{Outlier in the MIX}
\end{figure}

Figure \ref{Outlier in the MIX} shows the outlier in the \textbf{MIX} scenario. Initially, the blue surrounding vehicle is stopped due to the Driving Behavior Constraint defined in Section \ref{drc}, as the green ego vehicle is close. The ego vehicle attempts to accelerate and change lanes to overtake. The pink surrounding vehicle is also initially stopped, but after a while, it begins to change lanes at a low speed. Once the pink vehicle moves slightly, the driving behavior constraint on the blue vehicle is no longer triggered, allowing it to start moving. The abnormal behavior occurs when the pink vehicle suddenly stops, which appears unrealistic from the testers' perspective. Upon reviewing the log, it is clear that the pink vehicle triggered the side front action pattern, activating the brake condition. Since the pink vehicle was moving at a low speed, it could stop in a short time. Consequently, the ego vehicle failed to avoid the pink vehicle, resulting in a collision. This highlights the impact of our rule-based action pattern.

% \jz{you haven't explained why it is ranked not real, you explain this scenario is actually real, put some defences here to justify}

\subsubsection{Outlier Analysis II}
\begin{figure}[ht]
\centering
\raisebox{1mm}{\includegraphics[width=0.32\linewidth]{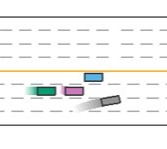}}
\raisebox{0.5mm}{\includegraphics[width=0.32\linewidth]{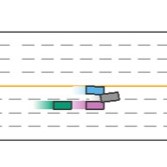}}
\includegraphics[width=0.32\linewidth]{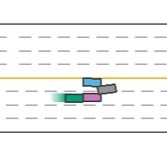}
\caption{Outlier in the straight scenario}
\label{Outlier in the straight}
\end{figure}
Figure \ref{Outlier in the straight} shows the outlier in the \textbf{Straight} scenario. The pink surrounding vehicle's driving behavior seems unrealistic, as it tries to intentionally block the green ego vehicle with a low speed. Even though another two surrounding vehicles collide around it, the pink vehicle continues to move at a low speed. Finally, the pink vehicle collides with those vehicles, and the ego vehicle collides with the pink vehicle. After checking the log, this could be explained by our Driving Behavior Constraint defined in Section \ref{drc}. As the two vehicles collided in front of the pink vehicle and stopped there, this triggered the driving behavior constraint of the pink vehicle to avoid colliding with other vehicles, causing it to move at a very low speed in an attempt to stop. However, due to its motion, it cannot stop completely, which results in the appearance of intentionally blocking the ego vehicle.

% \jz{this explanation is good, then you can mention as a future work we can relax this constraint using more advanced testing techniques which can select adaptively best possible next action while being feasible.}

\section{Discussion}
\label{dis}

In this work, we conduct an in-depth qualitative analysis of the latest simulator, MetaDrive. We use the built-in end-to-end driving models for evaluation: IDM, representing a rule-based model, and PPO, representing a learning-based model. Both are repetitive end-to-end learning models. Our approach is not limited to MetaDrive or end-to-end driving models alone; we plan to extend evaluation to other simulators, such as CARLA, and to multi-module driving systems, such as Baidu Apollo and Autoware.

Our primary focus is on detecting safety violations of the ego-vehicle in scenarios involving multiple surrounding vehicles, where the complex interplay among these vehicles can lead the ego-vehicle’s ADS into unsafe behaviors. This is a challenging task that state-of-the-art testing methods have failed to address, as shown in our experiments. Detailed records of these safety violations have been reported to the MetaDrive developers to help identify root causes. Early detection of potential safety issues is crucial for real-world ADS development and has the potential to be adapted to other machine learning-enabled cyber-physical systems, where dynamic objects in the environment can present unpredictable yet hazardous scenarios.
% This method can be applied early in the software development life cycle,  during the prototype and design stages, to detect bugs and guide design choices for machine learning models and system architecture. Early detection helps avoid costly fixes post-deployment. Using our simulation platform, we detected bugs and improved landing solutions, resulting in three generations of auto-landing systems with different machine learning models and architectures. Our method also aids in business operation risk assessment to identify potential risks and determine safe deployment areas, and detect early bugs in other critical components, such as the perception module.

% 
\section{Conclusion}
\label{conclusion}
In this paper, we present an innovative online testing method for generating safety violations involving multiple vehicles in ADS. Our approach combines a highly efficient multi-agent reinforcement learning (MARL) with a rule-based online fuzzer to produce realistic yet hazardous scenarios for the ego vehicle created by surrounding vehicles. Experiments show that \tool\ detects up to 136.2\% more safety violations than state-of-the-art methods. A user study further validates the realism of the generated scenarios. Future research will expand \tool\ by replacing the rule-based fuzzer with a more efficient online fuzzer, allowing \tool\ to explore the confined search space using MARL without rule constraints. We also plan to apply \tool\ to other machine learning-enabled cyber-physical systems, such as autonomous UAV landing systems.
%In this paper, we present an innovative method for generating safety violations that involve multiple vehicles in ADS. Our approach combines a fast-converging MARL and a rule-based online fuzzer to produce natural and common driving behaviors for surrounding vehicles. Experiments show that \tool\ detects up to 136.2\% more safety violations than the SOTA method. With the user study, we also validate the realism of our generated safety violation cases. These results demonstrate \tool's effectiveness in early error detection, offering significant safety and cost benefits. Future research will expand \tool\ with an efficient scenario generator to create more realistic corner cases, enabling more unique behaviors and broadening test scenarios. We also plan to apply \tool\ to other cyber-physical systems, such as UAV landing systems or robotics.

% we can relax this constraint using more advanced testing techniques which can select adaptively best possible next action while being feasible.

\bibliographystyle{ACM-Reference-Format}
\bibliography{reference}

% \appendix









\end{document}